\documentclass[journal]{IEEEtran}
\renewcommand\IEEEkeywordsname{Index Terms}
\IEEEoverridecommandlockouts
\usepackage{graphicx}
\usepackage[caption=false,font=footnotesize]{subfig}
\usepackage{amsmath}
\usepackage{epstopdf}
\usepackage{float}
\usepackage{color}
\usepackage{amssymb}
\usepackage{algorithm}
\usepackage{algpseudocode}
\usepackage{setspace}
\usepackage{booktabs} 
\usepackage{siunitx}  

\usepackage{makecell}

\usepackage{pifont}
\usepackage{graphicx}
\usepackage{tabularx}
\usepackage{booktabs}
\usepackage{adjustbox}
\usepackage{amsmath}
\usepackage{amssymb}
\usepackage{rotating}
\usepackage{multirow}
\usepackage{tikz}
\usepackage{xcolor}
\usepackage{xspace}
\usepackage{float}
\usepackage{comment}
\usepackage[normalem]{ulem}
\usepackage{tikz}
\usepackage{booktabs}
\usepackage{colortbl}
\usepackage{balance}
\definecolor{lightred}{RGB}{255, 204, 204}
\definecolor{lightyellow}{RGB}{255, 255, 153}
\definecolor{lightgreen}{RGB}{204, 255, 153}
\newcommand{\uparrowgreen}[1]{\tikz{\draw[blue,->,>=stealth,line width=0.02mm](0,0) -- (0,0.15);}\,\textcolor{blue}{\tiny #1\%}}
\newcommand{\downarrowred}[1]{\tikz{\draw[blue,->,>=stealth,line width=0.02mm](0,0.15) -- (0,0);}\,\textcolor{blue}{\tiny #1\%}}

\newcommand{\sol}{{{UniNet}}\xspace}
\newcommand{\rep}{{{T-Matrix}}\xspace}
\newcommand{\model}{{T-Attent}\xspace}

\newcommand{\hlt}[1]{{#1}}

\begin{document}

\title{\sol: A Unified Multi-granular Traffic Modeling Framework for Network Security}

\author{
		\IEEEauthorblockN{Binghui Wu, Dinil Mon Divakaran, and Mohan Gurusamy\\
}

        \IEEEcompsocitemizethanks{\IEEEcompsocthanksitem This article has been accepted for publication at IEEE Transactions on Cognitive Communications and Networking (2025).}
        
        \IEEEcompsocitemizethanks{\IEEEcompsocthanksitem Binghui Wu (\textit{Student Member, IEEE}) and Mohan Gurusamy (\textit{Senior Member, IEEE}) are with the Department of Electrical and Computer Engineering, National University of Singapore (NUS), 4 Engineering Drive 3, Singapore 117576. (email:binghuiwu@u.nus.edu, gmohan@nus.edu.sg).}
        \IEEEcompsocitemizethanks{\IEEEcompsocthanksitem Dinil Mon Divakaran (\textit{Senior Member, IEEE}) is with Institute for Infocomm Research (I$^2$R), A*STAR, Singapore (email: dinil\_divakaran@i2r.a-star.edu.sg).}
	}
\maketitle

\begin{abstract}
As modern networks grow increasingly complex---driven by diverse devices, encrypted protocols, and evolving threats---network traffic analysis has become critically important. Existing machine learning models often rely only on a single representation of packets or flows, limiting their ability to capture the contextual relationships essential for robust analysis. Furthermore, task-specific architectures for supervised, semi-supervised, and unsupervised learning lead to inefficiencies in adapting to varying data formats and security tasks. 

To address these gaps, we propose \sol, a unified framework that introduces a novel multi-granular traffic representation (\rep) with rich contextual information, integrating session, flow, and packet-level features to provide comprehensive contextual information. Combined with \model, a specially designed lightweight attention-based model, \sol efficiently learns latent embeddings for diverse security tasks. Extensive evaluations across four key network security and privacy problems---anomaly detection, attack classification, IoT device identification, and encrypted website fingerprinting---demonstrate \sol's significant performance gain over state-of-the-art methods, achieving higher accuracy, lower false positive rates, and improved scalability across various datasets. By addressing the limitations of single-level models and unifying traffic analysis paradigms, \sol sets a new benchmark for modern network security.

\end{abstract}

	
\section{Introduction}
\IEEEPARstart{O}{ver} the years, computer networks have evolved significantly due to the increase in network bandwidth, sophisticated network nodes (such as programmable switches), new device types (e.g., Internet of Things), changing network protocols (e.g., DNS-over-HTTPS), new applications (e.g., ChatGPT), etc. With this evolution also comes the challenge of securing the networks from various threats and attacks. 
Traditional rule-based systems have limitations in catching up with new and unknown threats; moreover, payloads are not available for deep packet inspection due to the increasing adoption of TLS~\cite{Google-transparency}. Consequently, researchers have long been exploring models from the domain of statistics, data mining, and machine learning (ML) to address the challenges in network traffic analysis~\cite{TLS-based-ML-2017, NDSS-2020-flowprint, evidence-gathering-2017,NADA-2018, GEE-2019, anomaly-detection-alpha-stable-2011, PCA-Diot-2004, Max-Entropy-Est-2005, DEFT-2019, ADEPT-2021}. The advancement in deep learning (DL) plays a crucial role in network traffic analysis for security tasks. These models leverage the vast and complex features of network traffic to identify anomalies and threats effectively. Additionally, with the advent of programmable switches~\cite{p4-original-2014}, there is potential for ML or partial ML logic to run directly on switches at terabits per second (Tbps) line rates~\cite{2021-CCS-programmable-switch, Unsenix2023-IDP-programmable-switch-flow}, promising real-time security capabilities. The deep learning models, from convolutional neural networks~(CNNs) to autoencoders and the latest transformer models~\cite{vaswani2017attention} are able to learn from large datasets consisting of hundreds of features. This has led to the development of several deep learning models for network anomaly detection, botnet detection, attack classification, fingerprinting and counter-fingerprinting of IoT devices and websites, traffic generation, and so on~\cite{KitSUne-2018-NDSS, Deepcorr-2018-CCS,botnet-S&P-2020, sigcom-2022-traffic-generation, doh-2021, Deepcoffea-2022-sp,shenoi2023ipet, Usenix-2023-website-fingerprinting, WU-ZEST,Netdifussion-2024-traffic-generation}.

\begin{figure*}[!htbp]
    \centering
    \includegraphics[width=0.85\linewidth]{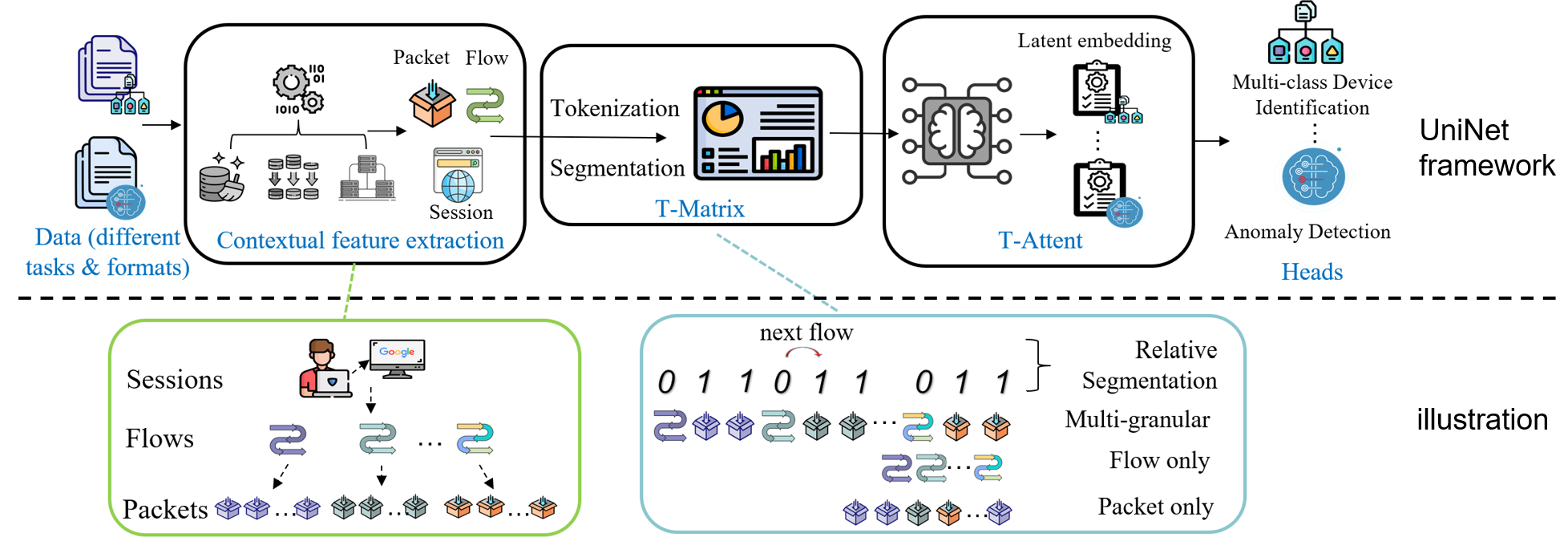} 
    \caption{Overview of \sol framework}
    \label{fig:overview-framework}
\end{figure*}

Despite these promising directions, a core challenge lies in data representation and formats. The common formats for network data are: i)~\texttt{pcap} that captures every packet on the wire and details from the headers ii)~\text{flows} (e.g., NetFlow, IPFIX~\cite{IPFIX}) that capture coarser information from an aggregation of packets. Packet captures provide rich details but require substantial resources to store and process; flow-based representations are more lightweight but lose important per-packet granularity. As a result, ML models must adapt to different levels of detail and data availability. Traditional intrusion detection systems (IDS) often focus on flows only, treating each flow as an isolated unit~\cite{disclosure-2012, AssoRuleMining-2012}. However, malicious behaviors rarely manifest in any single flow or packet in isolation. A single flow generally lacks conclusive evidence, and a lone packet offers minimal context unless considered within a broader temporal and relational environment. Therefore, recent efforts are shifting toward session-level representations, wherein flows sharing common attributes (e.g., source or destination IP addresses) within a certain time window are grouped into sessions. Session-level analysis provides more context than flow-level or packet-level views alone. However, most research works focus exclusively on one granularity at a time, which can either overlook subtle patterns critical for detecting sophisticated threats or demand excessive computational resources, undermining scalability and real-time applicability. 

\hlt{Recent research works have shown the ability to parse packets at line-rates for security use cases, e.g., rule-based DDoS detection and mitigation~\cite{Usenix-DDos-programme-switch-2021, dida2020, 
Unsenix2023-IDP-programmable-switch-flow}. 
Yet, representing all packets (of a traffic session) in a model is challenging. 
Firstly, using full packet sequences as inputs to the model makes the model size too large to be trained efficiently. Additionally, longer input sequences increase the inference time and make the system more vulnerable to certain attacks, e.g., DoS attacks specifically targeting such models. {\ding{172}}~Therefore, a key gap we identify in this domain is the lack of {\em efficient} and {\em effective} representation that includes both packet-, flow- and session-level features. Without such a representation, models are not easily adaptable for deployment across various networks that may have different traffic-capturing techniques and constraints, resulting in different data formats. {\ding{173}}~Furthermore, another critical gap is the uncertainty around the type of model best suited to handle these challenges. A general model that can function effectively across different conditions is important, as it ensures consistent performance despite variability in available features. Such a model must be capable of integrating various learning paradigms, including supervised, semi-supervised, and unsupervised learning. {\ding{174}}~Beyond the need for a general representation and a unified model, there is also a challenge of dealing with limited data. In scenarios where data is scarce, the ability to extract meaningful information and maintain robust performance becomes important. Table~\ref{tab:comaprison-with-works} presents a comparison with recent traffic analysis models.}

\begin{table}[h]
\centering
\caption{\hlt{Extended comparison with recent network traffic analysis models.}}
\label{tab:comaprison-with-works}
\renewcommand{\arraystretch}{1.3}
\resizebox{0.90\columnwidth}{!}{%
\begin{tabular}{lccccc}
\toprule
\rowcolor{white}
\textbf{Model} & 
\textbf{Multi-granular input} & 
\textbf{Multi-task} & 
\textbf{Multi-learning paradigms}\\
\midrule

AutoWFP~\cite{NDSS-2018-wfp-lstm}        & \cellcolor{lightred} No                     & \cellcolor{lightred} No     & \cellcolor{lightred} No     \\
TMWF~\cite{2023-Transformer-WFP-baseline}           & \cellcolor{lightred} No                     & \cellcolor{lightred} No     & \cellcolor{lightred} No     \\
TDoQ~\cite{Levi-2024}           & \cellcolor{lightred} No                     & \cellcolor{lightred} No     & \cellcolor{lightyellow}Yes (Supervised, Semi-supervised)     \\
SANE~\cite{WU-ZEST}           & \cellcolor{lightred} No                     & \cellcolor{lightred} No     & \cellcolor{lightred} No    \\
GRU-tFP~\cite{Usenix-2023-GRU-CNN}        & \cellcolor{lightred} No                     & \cellcolor{lightgreen} Yes  & \cellcolor{lightyellow} Yes (Supervised, un-supervised)    \\
BiLSTM-iFP~\cite{IoT-biLSTM-2020}     & \cellcolor{lightred} No                     & \cellcolor{lightred} No     & \cellcolor{lightred} No     \\
MNTD~\cite{MIL-Mutli-granular-2024}    & \cellcolor{lightyellow} Yes (Packet + Time)                     & \cellcolor{lightred} No     & \cellcolor{lightred} No     \\
\midrule
\textbf{UniNet (Ours)} & \cellcolor{lightgreen} \textbf{Yes (Session + Flow + Packet)} & 
\cellcolor{lightgreen} \textbf{Yes} & 
\cellcolor{lightgreen} \textbf{Yes (Supervised, Semi-, Unsupervised)} \\
\bottomrule
\end{tabular}%
}
\end{table}

To address these limitations, we introduce \sol, a unified framework designed to integrate multi-granular representations and support a broad range of network traffic analysis tasks. Figure~\ref{fig:overview-framework} provides an overview of \sol, highlighting its three main components: i)~\rep, A multi-granular traffic representation that can integrate session, flow, and packet level information; ii)~\model, A unified, self-attention-based feature extraction model capable of capturing contextual patterns from diverse data inputs; and iii)~heads tailored to different learning paradigms, including supervised, semi-supervised, and unsupervised tasks. Unlike previous approaches that either focus on flows or packets in isolation, \sol leverages these granularities in a single architecture, ensuring both fine-grained context and scalability. At the same time, its flexible architecture supports a variety of security and privacy tasks, from anomaly detection and attack classification to device identification and website fingerprinting (see Table~\ref{NTA tasks}).

\begin{table}[ht]
\centering
\caption{Tasks we consider for network traffic analysis (see threat model description in Section~\ref{Section:performance-evalue} for further details)}
\resizebox{0.9\columnwidth}{!}{%
\begin{tabular}{lllc}
\toprule
\textbf{Tasks} & \textbf{Learning paradigm} & \textbf{{Granularity}} & \textbf{Task ID} \\
\midrule
Anomaly detection & one-class un/semi-supervised & session, flow, packet & 1 \\
Attack identification & binary/multi-class supervised & flow, packet & 2 \\
Device identification & multi-class supervised & session, flow, packet & 3 \\
Website fingerprinting & multi-class semi-supervised & session, flow, packet & 4 \\
\bottomrule
\end{tabular}%
}
\label{NTA tasks}
\end{table}

The following summarizes our contributions:

\begin{enumerate}
\item \textbf{\rep}:
We develop a multi-granular representation for network traffic that is suitable for multiple data formats and their combinations~(Sections~\ref{sec:t-mat-def}). We carry out comprehensive experiments to compare \rep with single-level representations; the results show that \rep captures more detailed traffic patterns, leading to improved performance in various traffic analysis tasks~(Section~\ref{case_study-3-iot}).

\item \textbf{\model for latent embedding learning}: We develop a unified attention-based architecture for network traffic analysis that captures contextual information and simplifies model selection (Section~\ref{sec:model architecture}). \model effectively handles supervised, semi-supervised, and unsupervised learning by employing different ``heads” (Section~\ref{classification heads}). This design greatly reduces the overhead of using separate models for each task, making \sol a powerful choice for diverse traffic analysis scenarios (Sections~\ref{Section:performance-evalue}. Additionally, we adopt a lightweight variant of the transformer encoder and a new segmentation strategy (Section~\ref{sec: segmentation-embedding}), with reduced attention heads and embedding dimensions, which ensures computational efficiency without compromising performance.

\item \textbf{Enhanced efficiency and performance}: We evaluate \sol on four common network security and privacy tasks spanning three ML categories (unsupervised anomaly detection, supervised classification of attacks and devices, and semi-supervised website fingerprinting), using multiple real-world datasets (Section~\ref{Section:performance-evalue}). \sol consistently outperforms existing baselines in terms of detection rates and related metrics. Furthermore, we highlight ability of \sol to discover intrinsic patterns from limited data (Section~\ref{case_study-2-intrution-detction}). The self-attention mechanism in \model shows significant advantages in extracting information from informative sequences compared to baselines. We publish our code base for supporting future research and reproducibility\footnote{Code is available at: \texttt{{https://github.com/Binghui99/UniNet}}.}.

\end{enumerate}

\section{\sol framework}

We present an overview of our proposal, \sol. As depicted in Fig~\ref{fig:overview-framework}, \sol operates in four key steps. {i)~}The first step involves extracting semantic features at multiple levels, such as packet, flow, and session, to retain rich contextual information and meaningful fields; subsequently we define a multi-granular cohesive traffic representation \rep (Section~\ref{sec:t-mat-def}). {ii)~}{In the second step,} the unified \rep representation is encoded into tokens for training the model. In Section~\ref{sec:encoding}, we define the vocabulary of tokens corresponding to important traffic features and describe the tokenization process. {iii)~}After encoding the \rep representation of traffic into tokens, they are provided as input into the self-attention model, \model, for representation learning. We propose a relative segmentation embedding in Section~\ref{sec:model architecture}, which allows the model to identify and aggregate features at different levels, enhancing its ability to learn meaningful representations from the data. The output of \model is a latent embedding that represents the understanding of the traffic. {iv)~}This latent embedding is general enough to be used for various tasks, which is achieved by feeding it into different task-specific heads, {as explained in Section~\ref{classification heads}}. These heads provide a flexible framework for multiple network traffic analysis tasks. 

\section{\rep design}
\label{sec:t-mat-def}

\begin{figure*}[!htbp]
    \centering
    \includegraphics[width=0.8\linewidth]{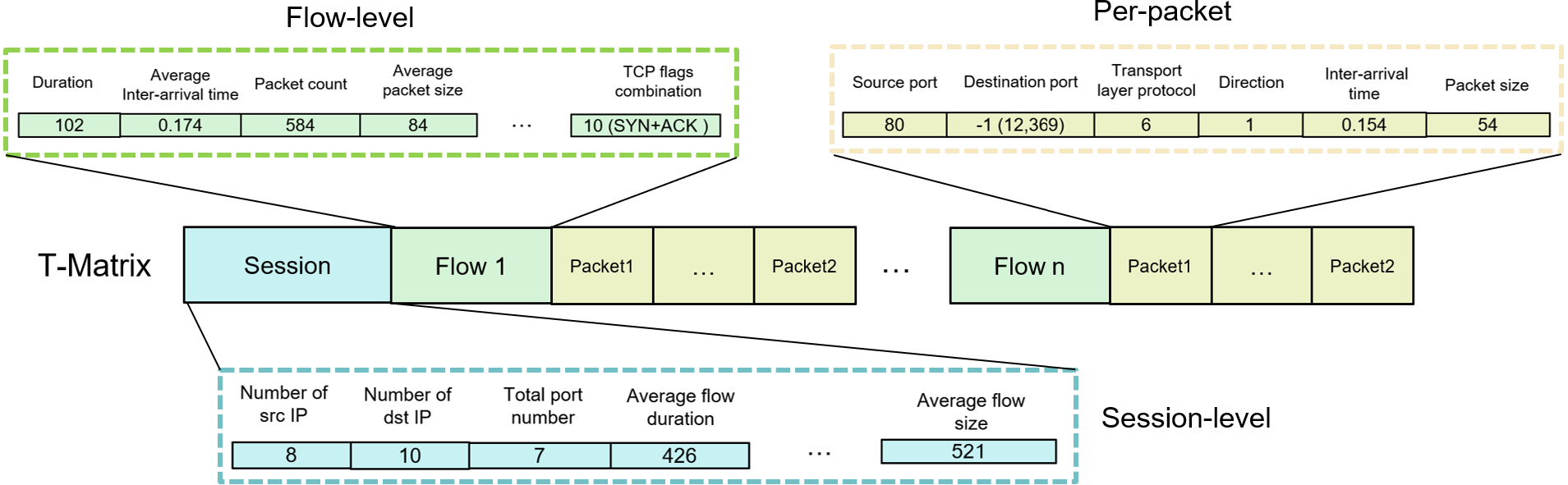} 
    \caption{\rep multi-granular traffic representation and defaulted session, flow, and packet level semantic features}
    \label{fig:T-matrixfeature}
\end{figure*}

\rep is a multi-granular traffic representation that encompasses information at three different levels of traffic information: session, flow, and packet. This is different from existing works that capture either flow-level or packet-level information but not both, thereby limiting the modeling capability. \hlt{Incorporating lightweight domain knowledge is often necessary to extract meaningful patterns from raw network traffic. \rep adopts basic yet generalizable features, such as port categories and TCP flag encodings, that are protocol-agnostic and widely validated in prior work~\cite{WU-ZEST, Levi-2024, IoT-biLSTM-2020, Usenix-2021-sequence-anomaly}. These features enable \sol to generalize across diverse tasks and protocols, without heavy reliance on manual feature engineering.
Traffic analysis systems typically operate by tapping traffic so that false positives (FPs), however low their number, do not interrupt normal connections. A stream of packets should be analyzed before decision-making. 
To support efficient analysis under this constraint, UniNet adopts a \textit{sliding window} strategy. Rather than waiting for a full traffic to complete, the system buffers packets within a fixed-size time window and extracts features from it.} We define {\em session} as a finite aggregation of {\em flows} that are temporally correlated and are contextualized by src/dst IP address. For example, a 15-minute traffic to and from one IP address forms a session. The separation of different sessions can be based on time (static) or based on inactivity (dynamic, e.g., `a silence of 1-min breaks a session into two'). Each {\em flow} in a session is a set of packets identified by the common 5-tuple of src/dst IP address, src/dst ports, and protocol. Thus, a session represents the behavior of, say, a user's browsing activity over a short period of time; the flows in the session describes the various connections, such as DNS query/response, HTTP request/response to different servers for various resources to load a website, and so on. Fig~\ref{fig:T-matrixfeature} illustrates the semantic multi-granular traffic representation of \rep.

Per-packet features are obtained from fields in the packet header. The raw packet features useful for traffic analysis include packet size, time since the last packet, packet direction, packet direction (incoming/outgoing), transport protocol (TCP/UDP), application protocol (HTTP, DNS, NTP, etc.), TLS presence and version, the categories of source/destination IP addresses (internal/external) and ports (service port, in particular). The port number helps to determine the type of application traffic, specifically differentiating between service (well-known) ports and ephemeral (random) ports. However, a single packet alone might not provide sufficient information for traffic analysis. Packet-level features are meaningful when a sequence of packets is considered. For example, a TCP SYN packet is present in both benign and malicious flow; as it does not {\em independently} help in determining whether the packet (and the corresponding flow) is malicious. However, when we analyze a sequence of packets, we may observe a rare pattern that indicates an anomaly; e.g., repetitive sequences with identical packet sizes, which are characteristic of application-layer DDoS attacks. Therefore, we extract these features from sequences of packets, encoding them (Section~\ref{sec:encoding}) to subsequently use the encoded features for training and inference. As payloads are (mostly) encrypted, we do not process payloads for feature extraction. 

Flow-level features are aggregated from the headers of  packets in a flow. This aggregation reduces the amount of data, but it is still useful when there are missing packet-level features due to resource limitations or when users tunnel through encrypted channels such as ToR and VPN. The identifier of a flow is the 5-tuple: src and dst IP addresses and port numbers, and transport protocol. Since data can flow in both directions, the forward and reverse flow identifiers are matched to learn the relationship. A silence period is used to determine the expiry of a 5-tuple flow within a session. There are tens of flow-level features that can be extracted from network traffic, and \sol is designed to represent a variable number of features. Some of the common flow-level features are flow size (in bytes and packets), flow duration, a combination of TCP flags, as well as statistical measures (mean, min, max, standard deviation, etc.) of sizes of all packets in the flow and inter-arrival times of packets, port numbers, and transport layer protocols~\cite{NDSS-2020-flowprint, ccs-2021-flow-level,NDSS-2021-flow-level-feature}. For clarity and systematic analysis, flag (e.g., ACK, SYN, FIN, PSH, URG, RST, ECE, CWR, and NS) combinations are numerically coded, which is illustrated as following:
    \begin{itemize}

    \item 1-9: Individual flags (e.g., ACK, SYN, FIN, PSH, URG, RST, ECE, CWR, NS).
    \item 10-14: Common combinations (SYN + AC = 10, PSH + ACK = 11, URG + ACK = 12, FIN + ACK = 13, RST + ACK = 14).
    \item 15: Reserved for any uncommon or previously unseen combinations.
    \end{itemize}
These features offer a balance between capturing essential characteristics and maintaining computational efficiency. Users have the flexibility to add or remove features as needed for their specific use cases.

At the session level, features provide information about the flows within. Consider a session aggregated using \textit{src} IP address (although it applies to other aggregations as well). This includes the total number o flows and \textit{dst} IP addresses, the unique number of \textit{dst} IP addresses, and the total number of service ports (e.g., 10 HTTP connections, 5 DNS resolutions). Such a representation allows us to detect some of the application-level anomalies, e.g., if there are 100s of outgoing DNS requests and no user application (such as browsing) in a short window, it might indicate an infected host. Given the above definition, \rep represents a session as a single data point. Since a session may consist of multiple flows, and each flow can contain multiple packets, flows and packets are represented as matrices. A session encapsulates aggregated information from its flows and is therefore represented as a single vector at the beginning of a data point.


\subsection{\rep Encoding}
\label{sec:encoding}

Next, we present the process of encoding the multi-granular semantic features extracted from traffic data into a standardized format suitable for \model, the second important component of \sol. The encoding process involves the following steps: tokenization, defining the vocabulary to represent features, and designing the final format for representing input.

\subsubsection{Tokenization}\label{tokenization}

Tokenization breaks down textual information into manageable units (tokens) that DL models can process and analyze~\cite{limisiewicz-etal-2023-tokenization}. All traffic features corresponding to a single data point (e.g., packet sequence) should be represented as a single token. In this way, the model provides insights into which specific features contributed to the detection of an anomaly, which not only enhances the ability to detect complex attack patterns but also improves the explainability of the results ({briefly discussed in Section~\ref{sec:discussion}). Unlike natural languages that share common characters and tokens, network traffic features are heterogeneous and the patterns are protocol-based~\cite{HotNets-rethinking}. As shown in Figure~\ref{fig:T-matrixfeature}, features such as direction, port number, protocol, and TCP flags are categorical, while packet length and inter-arrival time (IAT) are continuous. To unify this diverse data into a consistent format for model training, below we employ a tokenization method and define a vocabulary. Tokenization techniques~\cite{tokenization-metods-ACL-2023,iCASSP-tokenization-2024} split data into tokens. We handle categorical features by assigning each category a unique token, thereby converting data into a numerical format for processing. However, directly using continuous values can lead to poor model performance due to issues like overfitting and sensitivity to outliers~\cite{2019-continuous-bad-IJCAI,2021-CVPR-continous-bad,drawback-continuous-2021}. Therefore, we use binning to improve model convergence during training.

There are three commonly used binning methods~\cite{1995-bining-methods,bining-methods-nips-2022}: equal-width, equal-frequency, and clustering. Equal-width binning creates intervals of equal size, suitable for uniformly distributed data but it is less effective with outliers; in network traffic, attacks can be outliers. Equal-frequency binning distributes data points evenly across bins, managing skewed distributions well. Clustering, using algorithms like k-means, groups data by similarity, revealing inherent structures but requires more processing time~\cite{clustering}. We choose equal-frequency binning in \rep, for its efficiency and ability to minimize the impact of outliers.

\subsubsection{Vocabulary} 
\label{subsec:vocabular}

Vocabulary is the set of unique tokens a tokenization system utilizes during training. The design of the vocabulary must balance compression (using fewer tokens to represent more information) with model performance. While higher compression can speed up processing and extend context length, it may sacrifice the ability of models to capture fine-grained details~\cite{limisiewicz-etal-2023-tokenization}. A very small vocabulary size risks oversimplifying diverse data, leading to information loss and potential overfitting~\cite{vocabulary-size-2019-NAACL,vocabulary-2023}. On the other hand, a vocabulary that is too large can be computationally expensive and impractical given resource constraints~\cite{xlnet-2019, llama-2023}.

\begin{table}[ht]
\centering
\caption{Token IDs, Values, and Descriptions}
\resizebox{0.95\columnwidth}{!}{%
\begin{tabular}{>{\centering\arraybackslash}p{2cm}>{\centering\arraybackslash}p{1.7cm}>{\raggedright\arraybackslash}p{8.5cm}}
\toprule
\textbf{Token ID} & \textbf{Value} & \textbf{Description} \\ 
\toprule
0 & 0 & Token used for `\texttt{0}' in binary features. \\ 
1-1024 & 1-1024 & Conventional port numbers for specific services. \\
1025 & 8080 & HTTP port. \\ 
1026 & 3306 & MySQL port. \\ 
1027 & Other ports & Ports other than the specified well-known ports, and `\texttt{-1}' in port representations.\\ 
1028-1038 & Reserved & Reserved for future ports or protocols. \\ 
1040 & \texttt{[MASK]} & Masking purposes in representation learning. \\ 
1041 & \texttt{[PAD]} & Padding sequences in representation learning. \\
\bottomrule
\end{tabular}%
}

\label{tab:token list}
\end{table}

For categorical features, we need to decide the range of values. Port numbers are numerical identifiers used to distinguish different applications or services on a network, ranging from 0 to 65,535. However, using all 65,535 values is impractical, as it would require an immense amount of computational resources and result in large model sizes. Instead, we focus on commonly used ports that have significant meaning in traffic analysis. This includes the well-known ports from 1-1024, in addition to any custom application ports such as 8080 (\texttt{HTTP}) and 3306 (\texttt{MySQL}). 
Thus, we use 1024 as a base, adding specific tokens for special ports, future protocols, and other purposes. The final settings are given in Table~\ref{tab:token list}; the vocabulary size is 1042. This results in a total of 1042 tokens, including 2 special tokens, \texttt{[MASK]} and \texttt{[PAD]}, {for masked token prediction~(explained later in Section~\ref{mask-head})} and padding data with insufficient lengths. We bin continuous features into 1042 bins, which also function as normalization. As extreme values can impact this method, we carry out data cleaning to remove such values.

\subsection{\rep format}\label{final-format}

Considering the need to perform various network traffic analysis tasks, the input format must be sufficiently general to handle different scenarios. Thus, the input dataset will be in the format of a dictionary containing five keys:

\begin{enumerate}
    \item \texttt{input} represents the sequence generated in Section~\ref{tokenization}, containing information about the \texttt{[MASK]} token. The masking ratio $\mathbf{\eta}$ indicates the proportion of features in the tokenization that are masked. For example, when using the model for unsupervised learning tasks, such as anomaly detection, we set $ 0 < \mathbf{\eta} \leq 1$. For supervised classification tasks, we set $\mathbf{\eta} = 0$.
    \item \texttt{true value} represents the ground truth of the masked tokens. The values are all 0 except for the masked parts. To refine the loss function, we use the negative log loss function for model training.
    \item \texttt{mask index} indicates the indices of \texttt{[MASK]} tokens, facilitating the calculation of the loss function by identifying which parts of the input sequence are masked. 
    \item \texttt{segment label} separates session-level, flow-level, and packet-level features, indicating which features are at the flow level and which are at the packet level. We detect transitions between different flows by observing changes from $0 \rightarrow1$ or $1 \rightarrow 0$ in the segment label sequences.

    \item \texttt{sequence label} is used for handling supervised learning problems, providing labels for sequences to support classification and other tasks.
\end{enumerate}

An example is shown in Table~\ref{tab:input_dictionary}.

\begin{table}[htbp]
\centering
\caption{The illustration of final input format. The highlight parts represent the masked tokens.}
\resizebox{0.8\columnwidth}{!}{%
\begin{tabular}{ll}
\toprule
\textbf{Key} & \textbf{Example} \\
\midrule
\texttt{input} &  \texttt{[0,1,54,16,\textbf{1040,1040},5,1,1,...]} \\
\texttt{true value} & \texttt{[0,0,0,0,\textbf{45,85},1,1,...]} \\
\texttt{mask index}  & \texttt{[0,0,0,0,\textbf{1,1},0,0,...]} \\
\texttt{segment label}  & \texttt{[0,0,0,0,0,\textbf{0,0,1,1},...]} \\
\texttt{sequence label} & \texttt{[0] or [1] or [...]} \\
\bottomrule
\end{tabular}%
}
\label{tab:input_dictionary}
\end{table}

\section{\model Architecture}
\label{sec:model architecture}

\model is designed to handle the heterogeneous and diverse network traffic data by generating corresponding latent embeddings. The architecture of \model is shown in Figure~\ref{fig:model_arc}. It consists of several layers that work together to process and analyze the data effectively: embedding techniques, multiple encoder layers, and a masked prediction head for latent representation learning. 

\begin{figure}[htbp]
    \centering
    \includegraphics[width=1.0\linewidth]{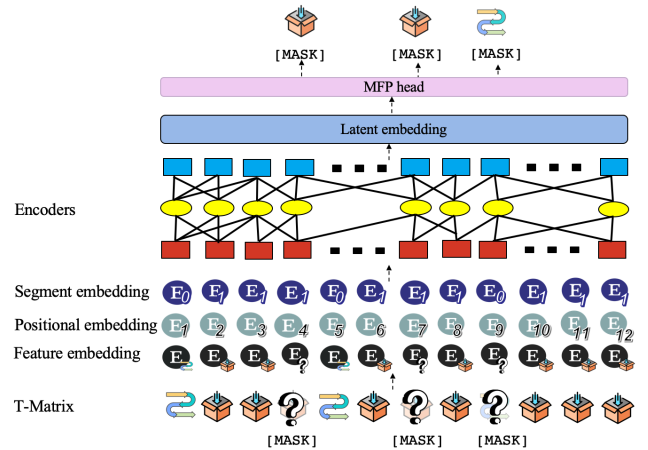} 
    \caption{The architecture of \model}
    \label{fig:model_arc}
\end{figure}

\subsection{Embedding and encoding layers}\label{sec: segmentation-embedding}

To effectively represent network traffic data within our attention-based model, \model employs several embedding techniques. 
\hlt{To enable the attention mechanism to effectively distinguish and integrate information across multiple granularities, we incorporate a \textit{hierarchical segmentation embedding}. Each input token is tagged with a segment label that identifies its level of granularity—specifically, packet-, flow-, or session-level. These labels are embedded alongside positional encodings and token embeddings, allowing the model to learn cross-level dependencies in a fully data-driven manner without imposing rigid attention constraints or handcrafted guidance.
For example, segment labels are assigned as follows: all packet-level tokens are labeled as \texttt{[0]}, flow-level tokens as \texttt{[1]}, and session-level tokens as \texttt{[2]}. A combined multi-granular input sequence may thus be represented by segment labels such as \texttt{[2, 1, 0, 0, 2, 1, 0, 0, \dots]}.
Each segment label is embedded into a learnable vector of the same dimension as the token and positional embeddings. If the embedding dimension is denoted as $d$, a segment label sequence of shape $1 \times N$ is mapped to a segment embedding matrix $S_{\text{emb}} \in \mathbb{R}^{N \times d}$. Likewise, positional embeddings are represented as $P_{\text{emb}} \in \mathbb{R}^{N \times d}$. The final input to the encoder is computed as the element-wise sum of the token embeddings $T_{\text{emb}}$, segment embeddings, and positional encodings: $$\text{Input} = T_{\text{emb}} + S_{\text{emb}} + P_{\text{emb}}.$$}
\hlt{This design enables the model to distinguish and attend across different granularities in a unified manner, while preserving architectural simplicity and maintaining generalizability across tasks.}
Additionally, the \model also leverages a lightweight ViT encoder layer~\cite{VIT} to process inputs from \rep. This encoder comprises a small number of attention heads and feed-forward layers. Unlike traditional transformers, \model uses a relative segmentation embedding mechanism to split inputs into multi-granular segments, enabling it to capture local structural patterns alongside high-level dependencies. The self-attention mechanism dynamically computes weights of different parts of the input sequence, allowing the model to capture interactions between packets and flows (e.g., linking a DNS lookup to a subsequent HTTP connection). Moreover, we utilize learnable positional embeddings~\cite{positional-embedding-iccv-2021, relative-positional-embedding-amazon} to encode the sequential order of packets within a flow, enabling the self-attention to capture essential temporal dependencies.

\subsection{\sol training with different heads}
\label{classification heads}

We now present the learning phase of our framework, where \sol is trained to generate encoded embeddings of network traffic data. These encoded embeddings can then be used as input for various ML heads or further processed for specific analysis tasks (as illustrated in Figure~\ref{fig:overview-framework}). We consider three heads for different purposes: unsupervised representation learning, anomaly detection, and classification. This framework allows \sol to be applied in different scenarios, enhancing its practical utility. 

\subsubsection{MFP head for unsupervised learning}\label{mask-head}

For unsupervised traffic representation learning (Section \ref{case_study_1 anomaly detection}), we introduce a new task called Masked Feature Prediction (MFP). This technique, inspired by the pretraining of LLMs~\cite{BERT}, involves intentionally masking certain tokens in the input data during training. The model is then trained to predict these masked tokens based on the surrounding context. For this purpose, we randomly select a percentage, denoted as $\eta$ (e.g., 40\%), of the features within a sequence to be masked. These selected features are replaced with the \texttt{[MASK]} token. The model is trained to predict the token IDs of these masked features using the provided ground truth values, used for unsupervised learning, as illustrated in Figure~\ref{fig:model_arc}.

\subsubsection{Anomaly detection head}\label{anomaly-head}

\hlt{The anomaly detection head is implemented as a lightweight autoencoder consisting of two fully connected layers in both the encoder and decoder. The encoder maps the latent embedding to a compressed bottleneck representation, followed by symmetric decoder layers to reconstruct the input. Formally, let $z$ denote the latent vector output from T-Attent. The encoder compresses $z$ as follows:
\[
h = \text{ReLU}(W_1 z + b_1), \quad \tilde{z} = \text{ReLU}(W_2 h + b_2)
\]
where $W_1, W_2 \in \mathbb{R}^{d \times d}$, and $d$ is the hidden size. The decoder mirrors this structure to produce the reconstruction $\hat{z}$. We compute the mean squared error (MSE) between $z$ and $\hat{z}$ as the reconstruction loss. Samples with losses exceeding the $\delta$-percentile threshold (e.g., 95th percentile of benign samples) are flagged as anomalous.}

\subsubsection{Classification head}~\label{calss-head}
\hlt{The classification head is a multi-layer perceptron (MLP) composed of two fully connected layers with ReLU activation, followed by a softmax output layer. Specifically, the MLP maps the latent embedding $z$ to a probability distribution over classes:
\[
h = \text{ReLU}(W_1 z + b_1), \quad y = \text{Softmax}(W_2 h + b_2)
\]
where $W_1 \in \mathbb{R}^{d \times d}$, $W_2 \in \mathbb{R}^{d \times C}$, and $C$ is the number of output classes. Cross-entropy loss is used for optimization in supervised tasks such as attack type identification or device classification.}

\section{Performance evaluations}\label{Section:performance-evalue}

\subsection{Experiments settings}\label{sec: Experiments-settings}

\hlt{We acknowledge the challenge posed by the limited availability of high-quality, open-source datasets~\cite{Badsmells-2024-eurosp}. To address this limitation, we intentionally selected three diverse datasets---CIC-IDS-2018~\cite{CIC-IDS20172018}, UNSW-2018~\cite{UNSW-2018}, and DoQ-2024~\cite{DoQ-2024-dataset}---collected by different institutions across various time periods (2018 to 2024), covering a wide range of protocols, attack types, and use cases. While CIC-IDS-2018 and UNSW-2018 were collected in controlled environments, the DoQ-2024 dataset includes real-world encrypted traffic captured during visits to live websites, based on modern web protocols such as HTTP/3 and DNS-over-QUIC, providing realistic noise and variability.}

\hlt{We evaluate our approach across four tasks. For anomaly detection and attack identification, benign traffic serves as background traffic, and the goal is to detect malicious flows hidden within normal activity. For IoT device classification, regular device communications represent typical network behavior, and the task is to correctly identify the device types. For website fingerprinting, particularly in the open-world setting, traffic from many unmonitored websites serves as background traffic, and the model must recognize specific monitored websites amid this noise.} All three datasets are extensive in both pcap and flow tabular formats, ensuring their suitability for our diverse tasks. Further details of each dataset are provided in the subsequent sections.

\begin{table}[htbp]
\centering
\caption{Default hyperparameters}
\resizebox{0.95\columnwidth}{!}{%
\begin{tabular}{lc}
\toprule
\textbf{Name}   & \textbf{Value}   \\

\midrule
Vocabulary Size & 1,042 \\

Number of Encoders & 2 \\

Embedding size & 10 \\

Batch Size & 32\\
Input length & 2,000 \\

Number of attention heads & 10 \\

Masking ratio & 40\% \\

Learning rate & 10e-4 warming up 10,000 steps  \\
Loss function & Negative Log Loss (Task 1), Cross-Entropy Loss (Tasks 2-4)\\


\bottomrule
\end{tabular}%
}

\label{tab:hyperparameters}
\end{table}


All the training and testing of our models and baselines are conducted on an Nvidia RTX~4080 16GB GPU and an Intel Core i9-13900KF processor. \hlt{Due to hardware constraints, we limited the embedding size and encoder depth to lightweight configurations, which also align with our goal of maintaining efficiency.  
We experiment with masking ratios ranging from 15\% to 60\%, finding optimal performance at 40\%. The vocabulary size for tokens is set to 1042. The model utilizes 10 heads, 10 embeddings, and 2 encoder layers. The learning rate follows a warm-up schedule, starting at 0.0001 and increasing to 0.001 over 10,000 steps. Specific settings for different heads are discussed in the corresponding sections. The default values are given in Table~\ref{tab:hyperparameters} for all tasks.} 

\begin{table}[htbp]
\centering
\caption{Binary confusion matrix.\text{TP/FP}:  True/False Positive; \text{TN/FN}: True/False Negative. }
\resizebox{0.8\columnwidth}{!}{%
\begin{tabular}{lcc}
\toprule
 & \textbf{Actual class: $Y$ } & \textbf{Actual class : not $Y$}\\
 \midrule
Predicted: $Y$ & \text{TP} & FP \\
Predicted: not $Y$ & \text{FN} & \text{TN} \\

\bottomrule
\end{tabular}%
}
\label{tab:confusion matrix}
\end{table}

\hlt{The commonly used metrics for network security tasks include Recall (True Positive Rate, TPR), Precision, False Positive Rate (FPR), Accuracy, and Area Under the Curve (AUC). AUC is a popular metric that counters the adverse effects of class imbalance. According to Table~\ref{tab:confusion matrix}, the metrics are calculated by equations:}

\[
\text{Recall} = \frac{\text{TP}}{\text{TP} + \text{FN}}, \quad
\text{Precision} = \frac{\text{TP}}{\text{TP} + \text{FP}}
\]
\[
\text{FPR} = \frac{\text{FP}}{\text{FP} + \text{TN}}, \quad
\text{Accuracy} = \frac{\text{TP} + \text{TN}}{\text{TP} + \text{TN} + \text{FP} + \text{FN}}
\]

For multi-class classification, we compute the \textit{macro} values of these metrics independently for each class and then average them across all classes. The threat model for each task is mentioned in the corresponding sections. We now evaluate \sol and baselines for four different security tasks---Tasks~1-4 in Table~\ref{NTA tasks}---across the three categories of unsupervised anomaly detection, supervised classification of attacks and devices, and semi-supervised website fingerprinting.

\subsection{\textbf{Task 1: Unsupervised Anomaly Detection}}\label{case_study_1 anomaly detection}

\noindent{\bf Threat model:} In anomaly detection (Task 1), the primary goal is to detect malicious network traffic that deviates from a learned benign profile. We assume that the training dataset, organized into session-level structures, is predominantly benign but may contain a small fraction of undiscovered attacks; however, it is not extensively poisoned by adversaries. Attackers can manipulate or inject flows, adjusting timing or header fields (e.g., IP addresses, ports) to blend into normal patterns; however, they do not control the overall training pipeline or the underlying network infrastructure.

\noindent{\bf Dataset:} We use the CSE-CIC-IDS2018 dataset~\cite{CIC-IDS20172018} for this task, and after processing the input into \rep format, we use only the benign traffic to train \model\footnote{In practice, the benign class is created by removing suspicious flows using rules; yet it is assumed that small part of this class contains some malicious flows~\cite{GEE-2019}}. The evaluation focuses on five types of network attacks: DDoS, DoS, BruteForce, Botnet, and Infiltration, which are categorized as malicious during the testing phase. The distribution of training and testing data is detailed in Table~\ref{tab:data_distribution of anomaly detection}.

\begin{table}[htbp]
\centering
\caption{Data distribution for anomaly detection (Task 1) }
\resizebox{0.9\columnwidth}{!}{%
\begin{tabular}{llrccc}
\toprule
\textbf{Category} & \textbf{Type} & \textbf{Count} & \textbf{Distribution~(\%)} & \textbf{Label} & \textbf{Ratio ~(\%)} \\
\midrule
\textbf{Training} & Benign   & \textbf{223,662} & -& - & -\\
\midrule
\midrule
\multirow{7}{*}{\textbf{Testing}} & Benign          & 10,000  & 50 & {0} & {50} \\ 
\midrule
 & DDoS           & 2,000  & 10 &  \multirow{6}{*}{1} & \multirow{6}{*}{50}\\ 
 & DoS            & 2,000   & 10 &  & \\
 & BruteForce     & 2,000    & 10 &  & \\
 & Bot            & 2,000    & 10 &  & \\
 & Infiltration   & 2,000   & 10 &  & \\
\bottomrule
\end{tabular}%
}

\label{tab:data_distribution of anomaly detection}
\end{table}

\noindent{\bf Input representation:} The input to \sol is structured to facilitate unsupervised learning, organized at a session level. Sessions are composed of flows grouped by the same source or destination IP (Section~\ref{sec:t-mat-def}). Segment labels distinguish different levels of features and different flows within the same session. The input sequence length is set to 2,000 tokens. We input all flows and their packets in the order of arrival until the sequence reaches 2,000 tokens. Any remaining tokens are padded with \texttt{[PAD]}. Each flow is represented by 8 features, and each packet by 6 features (Section~\ref{sec:t-mat-def}). Thus, representing a flow-packet segment requires a length of 68 features, making space for $\approx30$ flows within an input sequence. Given the simpler and less informative nature of packet features compared to natural language, a higher masking ratio is justified.

\begin{table*}[htbp]
    \centering   
    \caption{Comparison of Baseline Models and {\sol} for Task 1}
    \resizebox{0.8\textwidth}{!}{%
    \begin{tabular}{llcccccc}
        \toprule
        & \textbf{Model} & \textbf{Accuracy} & \textbf{F1 Score} & \textbf{Precision} & \textbf{Recall} & \textbf{AUC} & \textbf{FPR} \\
        \midrule
        \multirow{8}{*}{\textbf{Baseline}} 
        & Isolation Forest & 0.5260 & 0.5537 & 0.5299 & 0.5760 & 0.5312 & 0.3124 \\
        & One-Class SVM & 0.6412 & 0.6337 & 0.6220 & 0.6468 & 0.6490 & 0.2581 \\
        & LOF & 0.6918 & 0.6719 & 0.6505 & 0.6960 & 0.6907 & 0.2893 \\
        & K-means & 0.5804 & 0.5356 & 0.5831 & 0.5412 & 0.5798 & 0.4190 \\
        & AE & 0.6204 & 0.6037 & 0.6019 & 0.6275 & 0.6212 & 0.2750 \\
        & VAE & 0.7112 & 0.7156 & 0.6924 & 0.7405 & 0.7321 & 0.2645 \\
        & LSTM-VAE & 0.7351 & 0.7357 & 0.7348 & 0.7279 & 0.7660 & 0.2336 \\
        \cmidrule{2-8}
        & \textbf{Average} & 0.6437 & 0.6357 & 0.6306 & 0.6511 & 0.6528 & 0.2931 \\
        \midrule
        \multirow{8}{*}{\textbf{\sol+}} 
        & Isolation Forest head & 0.6427 {\scriptsize \uparrowgreen{22.16}} & 0.6594 {\scriptsize \uparrowgreen{19.16}} & 0.6487 {\scriptsize \uparrowgreen{22.18}} & 0.6815 {\scriptsize \uparrowgreen{18.06}} & 0.6728 {\scriptsize \uparrowgreen{26.41}} & 0.2825 {\scriptsize \downarrowred{9.56}} \\
        & One-Class SVM head& 0.7521 {\scriptsize \uparrowgreen{17.29}} & 0.7435 {\scriptsize \uparrowgreen{17.10}} & 0.7306 {\scriptsize \uparrowgreen{17.49}} & 0.7579 {\scriptsize \uparrowgreen{17.20}} & 0.7552 {\scriptsize \uparrowgreen{16.36}} & 0.2205 {\scriptsize \downarrowred{14.57}} \\
        & LOF head & 0.7814 {\scriptsize \uparrowgreen{12.95}} & 0.7698 {\scriptsize \uparrowgreen{14.58}} & 0.7612 {\scriptsize \uparrowgreen{17.01}} & 0.7807 {\scriptsize \uparrowgreen{12.16}} & 0.7793 {\scriptsize \uparrowgreen{12.81}} & 0.2618 {\scriptsize \downarrowred{9.52}} \\
        & K-means head & 0.6549 {\scriptsize \uparrowgreen{12.84}} & 0.6403 {\scriptsize \uparrowgreen{19.55}} & 0.6621 {\scriptsize \uparrowgreen{13.54}} & 0.6304 {\scriptsize \uparrowgreen{16.45}} & 0.6532 {\scriptsize \uparrowgreen{12.65}} & 0.3760 {\scriptsize \downarrowred{10.26}} \\
        & AE head & 0.7854 {\scriptsize \uparrowgreen{26.60}} & 0.7742 {\scriptsize \uparrowgreen{28.26}} & 0.7531 {\scriptsize \uparrowgreen{25.15}} & 0.7967 {\scriptsize \uparrowgreen{27.50}} & 0.7835 {\scriptsize \uparrowgreen{26.10}} & 0.2154 {\scriptsize \downarrowred{21.68}} \\
        & VAE head & 0.8023 {\scriptsize \uparrowgreen{12.84}} & 0.7927 {\scriptsize \uparrowgreen{10.77}} & 0.7709 {\scriptsize \uparrowgreen{11.34}} & 0.8163 {\scriptsize \uparrowgreen{10.24}} & 0.8034 {\scriptsize \uparrowgreen{9.73}} & 0.1968 {\scriptsize \downarrowred{25.59}} \\
        & LSTM-VAE head & 0.8689 {\scriptsize \uparrowgreen{13.57}} & 0.8584 {\scriptsize \uparrowgreen{13.61}} & 0.8497 {\scriptsize \uparrowgreen{15.64}} & 0.8679 {\scriptsize \uparrowgreen{11.58}} & 0.8681 {\scriptsize \uparrowgreen{13.37}} & 0.1312 {\scriptsize \downarrowred{43.81}} \\
        \cmidrule{2-8}
        & \textbf{Average} & 0.7597 {\scriptsize \uparrowgreen{18.01}} & 0.7526 {\scriptsize \uparrowgreen{18.49}} & 0.7438 {\scriptsize \uparrowgreen{17.98}} & 0.7659 {\scriptsize \uparrowgreen{17.64}} & 0.7637 {\scriptsize \uparrowgreen{17.00}} & 0.2406 {\scriptsize \downarrowred{17.90}} \\
        \bottomrule
    \end{tabular}%
    }

    \label{tab:comparison}
\end{table*}


\noindent{\bf Baselines:}  The baselines we evaluate are:
\begin{enumerate}
    \item \textbf{Machine Learning baselines:} We consider traditional ML algorithms such as Isolation Forest, One-Class SVM, Local Outlier Factor (LOF), and K-means clustering. These models rely on statistical and distance-based methods to identify anomalies. They are particularly effective for scenarios with well-defined feature spaces, offering faster training times and lower computational requirements. They have been used commonly for network traffic analysis (e.g., see~\cite{isolation-forest-2018,one-classSVM-2003,LOF-2019,kmeans-2008-anomaly}).
    
    \item \textbf{Deep learning baselines:} We implement deep learning models {used in the past for network anomaly detection}, including standard autoencoders (AE)~\cite{GEE-2019}, variational autoencoders (VAE)~\cite{VAE-2018}, and LSTM-based VAEs~\cite{LSTM-VAE}. These models are good at learning hierarchical and temporal representations from raw network traffic data. AE reconstructs input data and detects anomalies based on reconstruction loss, while VAEs introduce a probabilistic framework to model data distributions. LSTM-based VAEs capture sequential dependencies in traffic patterns, enhancing anomaly detection for time-series data. 
\end{enumerate}

The primary distinction between \sol and the baseline approaches lies in the utilization of the MFP head for embedding extraction and multi-granular representation. Specifically, \sol employs \rep and embeddings generated by the MFP head, which are subsequently processed through various anomaly detection models. In contrast, baseline approaches use single-level information, such as a sequence of packets or flows. They skip this step and apply anomaly detection techniques directly to features without encoding by the MFP head.

\noindent{\bf Orchestration of \sol:} To address these threats, we employ \sol in a two-phase, unsupervised fashion. Firstly, the MFP head (Section~\ref{mask-head}) learns representative embeddings by randomly masking up to 40\% of traffic features and predicting them, enabling the model to capture robust patterns of benign behavior. Once \model training is complete, the MFP head is removed, and the latent embeddings generated by the final encoder layer is utilized in the next phase. Secondly, an autoencoder-based anomaly detection head refines these embeddings, using reconstruction loss to identify deviations from the learned profile.


\noindent{\bf Analysis:} We present the performance of each model, both for the baselines and for our enhanced implementation using \sol~(i.e., \sol + different heads) in Table~\ref{tab:comparison}. For \sol, we perform unsupervised representation learning using the MFP head (Section~\ref{mask-head}) with \model. Subsequently, the initial traffic data is embedded into a transformed space to better capture underlying patterns and anomalies. The embeddings generated by \model are then fed into different baseline models (anomaly detection heads). 

As depicted in Table~\ref{tab:comparison}, \sol consistently outperforms the baseline across all key metrics --- the accuracy improves by an average of 18.01\%, F1-score by 18.49\%, precision by 17.98\%, recall by 17.64\%, and AUC by 17.00\%. The enhancements are even more pronounced with deep learning models; in comparison to AE, \sol registers a maximum improvement of (approximately) 27\% in accuracy, 28\% in F1-score, and a reduction of about 44\% in FPR. These results show that {\sol} accurately detects anomalous traffic patterns while significantly reducing the false positive rates. The enhanced performance of {\sol} can be attributed to the effective representation learning capabilities of the MFP head when combined with \model. The embedding generated by \model encompasses both sequential and statistical features, leading to a more robust and comprehensive understanding.

\subsection{\textbf{Task 2: Supervised Attack Identification}}\label{case_study-2-intrution-detction}

\noindent{\bf Threat model:} As for attack identification (Task~2), we consider a realistic network environment where attackers launch a variety of threats, while attempting to evade detection by mimicking benign traffic patterns and manipulating both flow- and session-level characteristics. An IDS aims first to distinguish malicious from benign traffic (Task~2.1), using a coarse-grained yet efficient binary classifier to handle high volumes of data. Flows flagged as malicious are then subjected to a second, more detailed classification step (Task~2.2), which identifies the specific attack type (e.g., botnet, DDoS) using a multi-class head that requires deeper contextual analysis. 

A significant challenge inherent in this environment is the scarcity of labeled instances for training. Attackers often exploit this weakness, as obtaining large numbers of labeled samples for diverse or emerging attack types is prohibitively costly and time-consuming in real-world settings. This lack of labeled data can hinder the IDS’s ability to generalize to new threats or achieve high classification accuracy.

\noindent{\bf Dataset:} We utilize the CSE-CIC-IDS2018 dataset~\cite{CIC-IDS20172018}, which is predominantly composed of benign samples, reflecting real-world class imbalances and the limited availability of labeled data for certain attack types. We focus on four types of attacks: DoS, brute force, botnet, and infiltration. In Phase 1, all attacks are aggregated into a single malicious class. Phase 2 refines this classification by distinguishing among individual attack types. To address data imbalance, additional preprocessing steps are applied. The data distribution for Task~2 is presented in Table~\ref{tab:balanced_intrusion_distribution}.

\begin{table}[htbp]
\centering
\caption{Intrusion detection data distribution (Task 2)}
\resizebox{0.9\columnwidth}{!}{%
\begin{tabular}{llrrccc}
\toprule
\multirow{2}{*}{\textbf{Category}} & \multirow{2}{*}{\textbf{Type}} & \multirow{2}{*}{\textbf{Count}} & \multirow{2}{*}{\textbf{Ratio~(\%)}} & \multicolumn{2}{c}{\textbf{Labels}} & \multirow{2}{*}{\textbf{Total Ratio ~(\%)}} \\ 
 & & & & {Phase 1} & {Phase 2} & \\
\midrule
\textbf{Benign} & Benign  & \textbf{40,000} & 57.04 & 0 & 0 & 57.04 \\
\midrule
\multirow{4}{*}{\textbf{Attack}} & DoS          & 10,196  & 14.54 & \multirow{4}{*}{1} & 1  & \multirow{4}{*}{42.96} \\
                                 & BruteForce   & 9,523   & 13.58 &                      & 2  & \\
                                 & Bot          & 6,359   & 9.07  &                      & 3  & \\
                                 & Infiltration & 4,048   & 5.77  &                      & 4  & \\
\bottomrule
\end{tabular}%
}

\label{tab:balanced_intrusion_distribution}
\end{table}

\noindent{\bf Input format:} In this task, the classification is based on a single flow. It focuses on classifying individual flows based on flow-level statistics and short-term packet patterns, which is more localized in nature. Therefore, an input is a single flow and set of packets within the flow, with a length of 2,000 tokens. This format begins with flow-level features, followed by packet-level features within the same flow. If the number of packets in a flow exceeds the maximum length of the input, it will be truncated. And if the number of packets is less than the fixed length, it will be padded with \texttt{[PAD]}. We use the default flow and packet features described in Section~\ref{sec:t-mat-def}.
The segment labels are used to {separate per-packet and flow-level features, indicating which level a particular feature belongs to.}

\begin{figure*}[htbp]
    \centering
    \subfloat[Performance of models for Task~2.1]{%
        \includegraphics[width=0.32\textwidth]{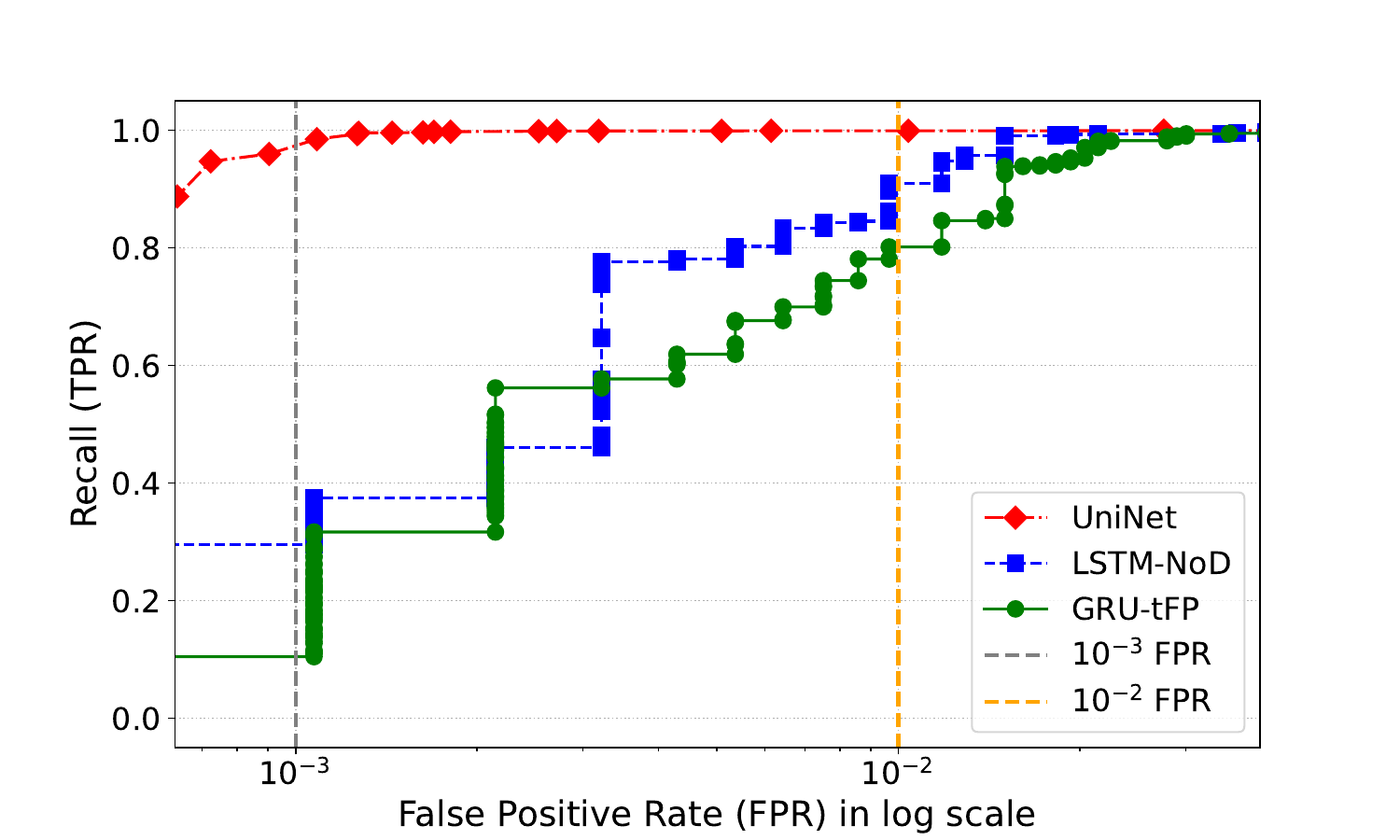}%
        \label{fig:task2-ROC}
    }
    \hfill
    \subfloat[Performance of models for Task~2.2]{%
        \includegraphics[width=0.32\textwidth]{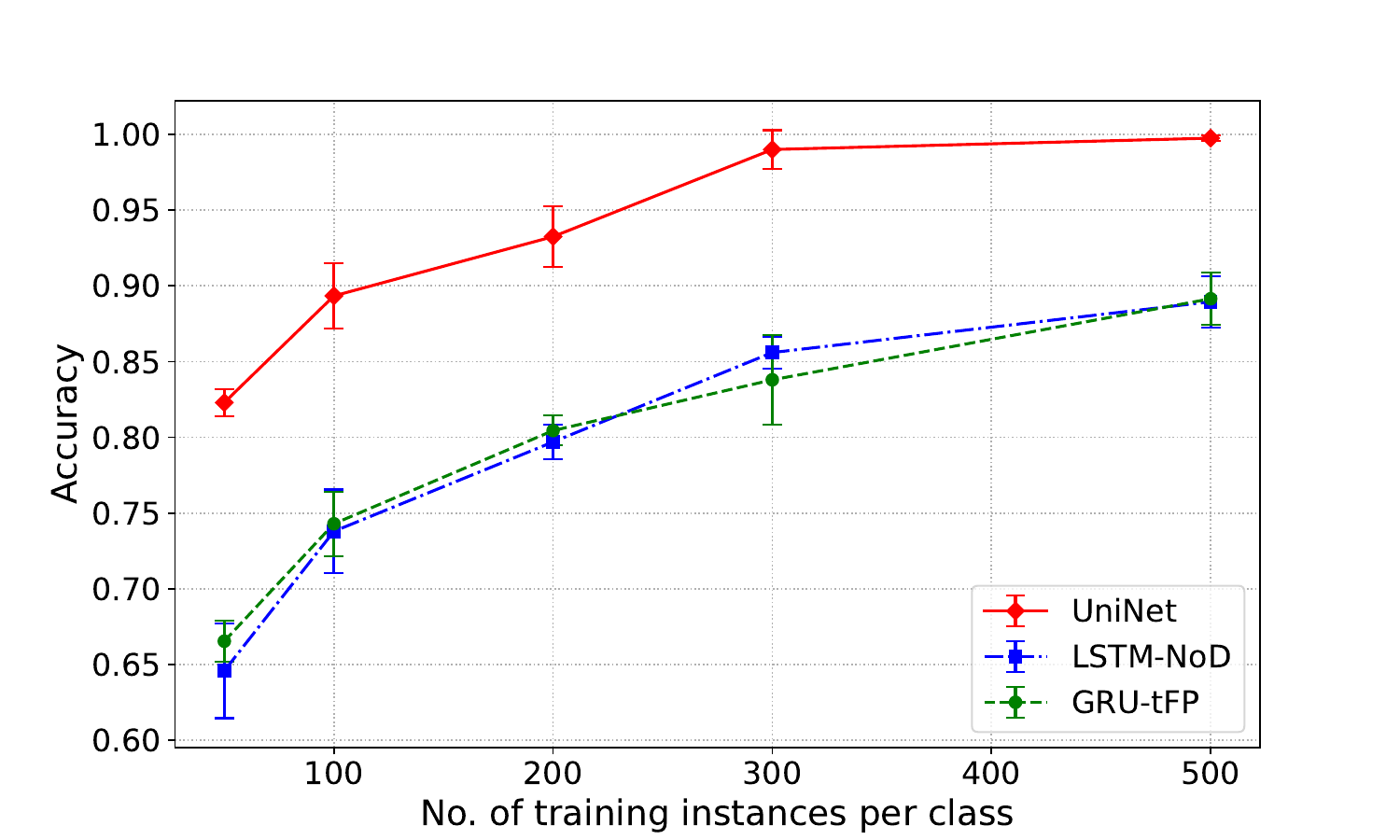}%
        \label{fig:overall-accuracy}
    }
    \hfill
    \subfloat[F1-score of DoS]{%
        \includegraphics[width=0.32\textwidth]{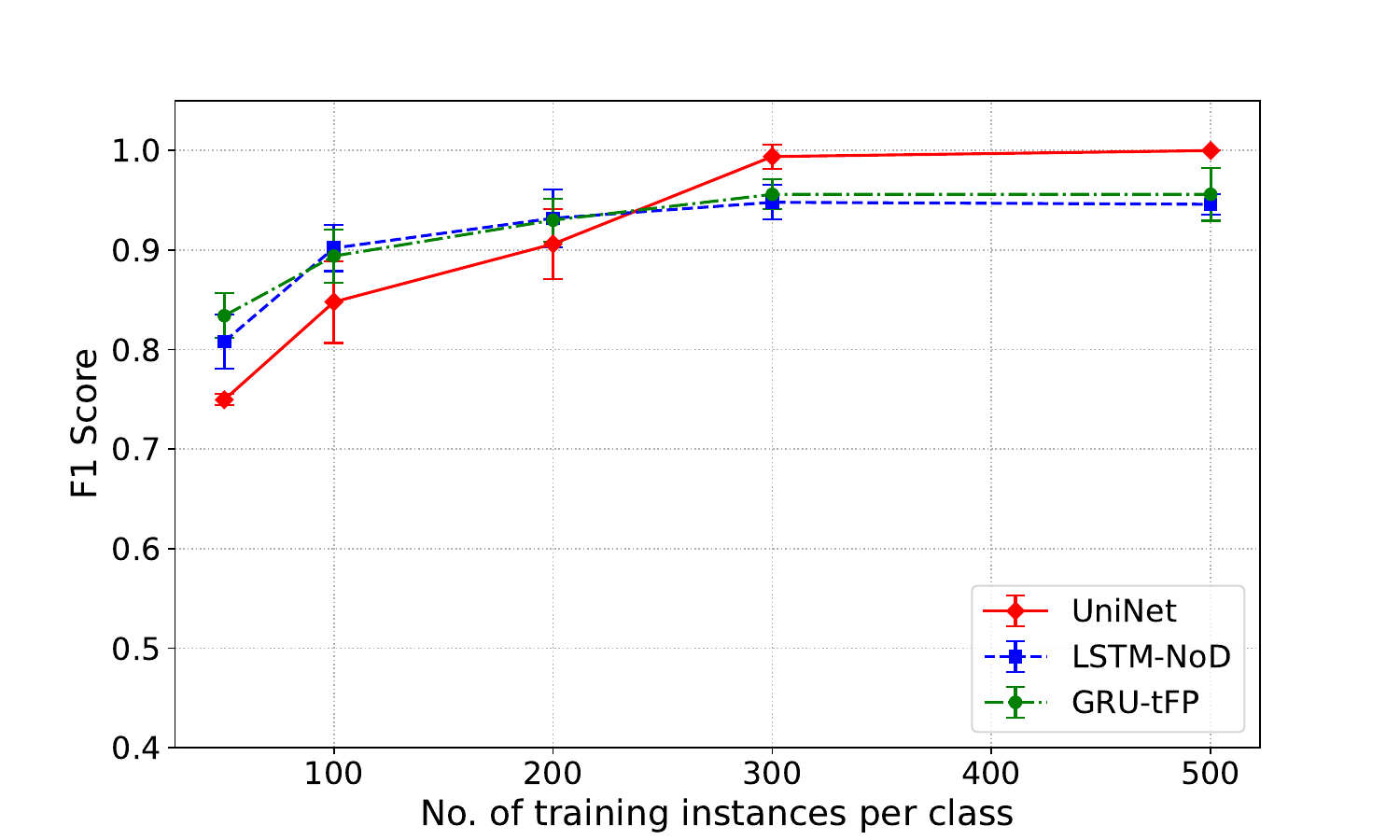}%
        \label{fig:dos}
    }
    \hfill
    \subfloat[F1-score of Bruteforce]{%
        \includegraphics[width=0.32\textwidth]{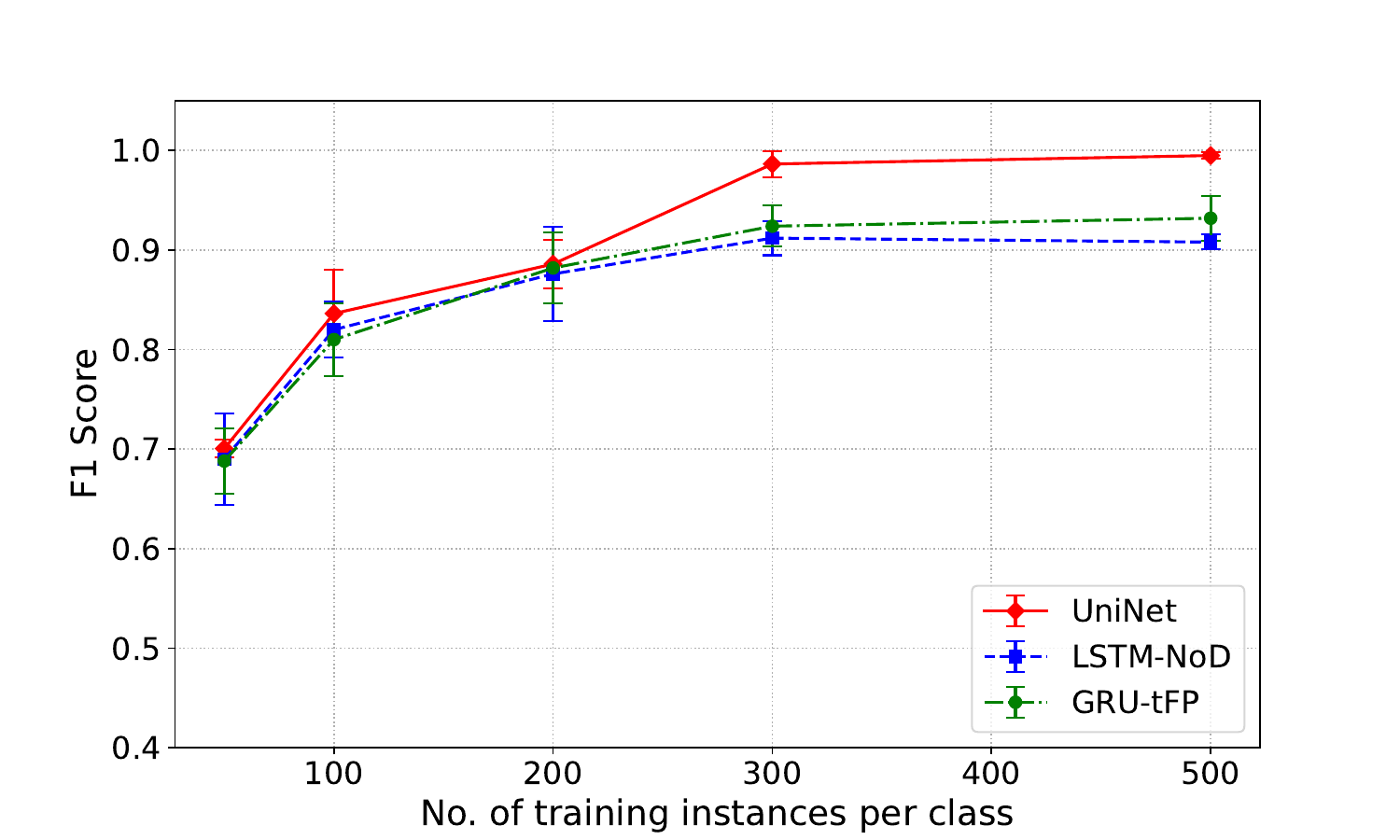}%
        \label{fig:bruteforce}
    }
    \hfill
    \subfloat[F1-score of Infiltration]{%
        \includegraphics[width=0.32\textwidth]{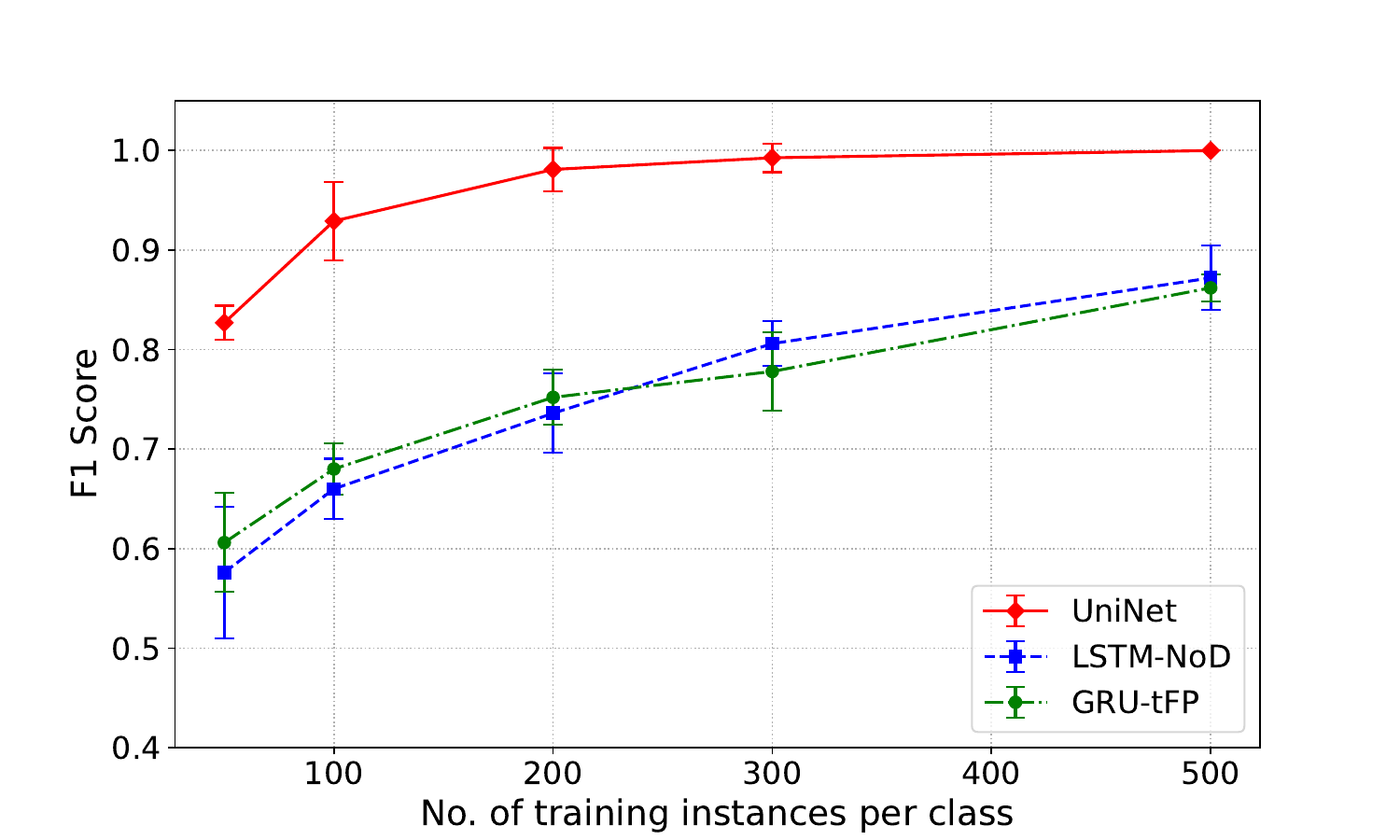}%
        \label{fig:infiltration}
    }
    \hfill
    \subfloat[F1-score of Botnet]{%
        \includegraphics[width=0.32\textwidth]{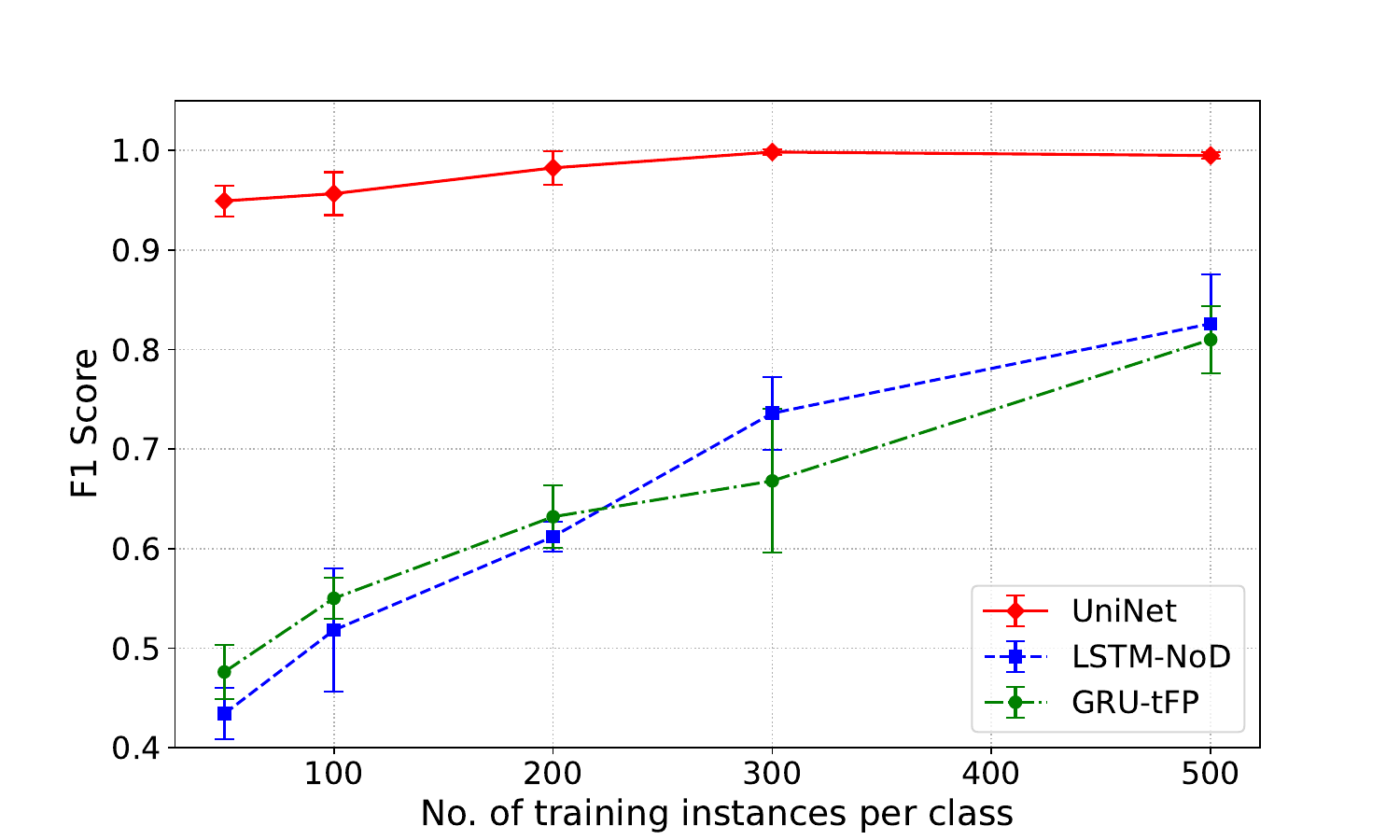}%
        \label{fig:botnet}
    }

    \caption{F1-scores and performance metrics for various attack types and phases in Task 2.}
\end{figure*}

\noindent{\bf Baselines:} We compare \sol with recent sequence models: LSTM-NoD~\cite{AsicaCCS-2019-LSTM-model} and GRU-tFP~(Gated Recurrent Unit)~\cite{Usenix-2023-GRU-CNN}. The LSTM-NoD model utilizes two LSTM models, one trained on normal-day (N) traffic and the other on attack-day (D) traffic, to estimate the likelihood of network requests being DDoS attacks~\cite{AsicaCCS-2019-LSTM-model}. The GRU-tFP model is introduced in~\cite{Usenix-2023-GRU-CNN} to address different tasks, including intrusion detection, in a supervised way. GRU-tFP uses the GRU model to extract traffic features hierarchically to capture both intra-flow and inter-flow correlations. To analyze the impact of \rep and \model, LSTM-NoD and GRU-tFP are provided with single packet-level data sequences. In contrast, \sol uses the \rep format with flow and packet level data. We also assess the ability of each model to extract meaningful traffic patterns with limited training instances per class. We employ a lightweight hierarchical transformer architecture comprising two encoder layers with an embedding size of 10, resulting in a total of 15,000 parameters. This parameter count is significantly smaller compared to LLMs, which typically contain billions of parameters. The compact design facilitates efficient execution and simplifies implementation, making it suitable for deployment in resource-constrained environments.

\noindent{\bf Orchestration of \sol:} While adversaries may manipulate timing and header fields to blend in with legitimate sessions, the IDS leverages session-level aggregation, flow-based features, and specialized embedding strategies to highlight anomalies that cannot be entirely concealed. Under conditions of label sparsity, we design experiments that explore the system's robustness under varying levels of labeled data availability, ranging from highly sparse (50 samples per class) to more representative distributions (500 samples per class).

\noindent{\bf Analysis:} For the Phase 1 (broad detection), the results are presented in Table~\ref{tab:performance_scores_task2}. {\sol} achieves the highest accuracy of 99.41\% over all baselines. In the context of intrusion detection, balancing the trade-off between recall/TPR (True Positive Rate) and the False Positive Rate (FPR) is crucial. A low FPR is essential to minimize false alarms, which cost human hours for security analysis. However, this often comes at the expense of recall, due to missed detection of anomalies. Figure~\ref{fig:task2-ROC} illustrates the performance of \sol and baseline models across different FPR values. All models achieve high recall at high FPR levels, but the real test of efficacy lies in their performance at lower FPR values. At an FPR of $10^{-2}$, \sol demonstrates an absolute increase of $\sim 14\%$ for TPR compared to the best-performing baseline (LSTM-NoD). This advantage becomes even more notable as the FPR is reduced to $10^{-3}$; the TPR gap between \sol and best performing baseline increases significantly to $\sim 68\%$. These results highlight the ability of \sol to maintain high detection rates without sacrificing the FPR.

We test with different \textit{training instances per class} to evaluate the information extraction capability of different models for Phase 2 (granular classification). When provided with same informative data, the model that extracts and utilizes information most effectively has a significant impact. Figure~\ref{fig:overall-accuracy} gives the overall accuracy across all attacks, where \sol exhibits an average $\sim 14\%$ accuracy improvement over the baselines. 
The model converges with 300 training instances per class, highlighting the effectiveness of \model part in \sol, which utilizes the self-attention mechanism to extract intrinsic patterns.

Figure~\ref{fig:dos}-\ref{fig:botnet} shows the F1-scores for each attack type. Although DoS and Brute Force attacks are generally easier for all models to detect due to their prominent and distinguishable characteristics, we still see an increasing gap between \sol and baselines with increasing training instances.
As for Bot and Infiltration attacks, {\sol demonstrates a significant improvement over LSTM-NoD and GRU-tFP, particularly with a low number of training instances (e.g., 100) per class. Notably, there is an absolute increase in F1-score by $\sim 25\%$ for Infiltration and $\sim 43\%$ for Botnet compared to the best-performing baseline (GRU-tFP). This can be attributed to the limitations of LSTM-NoD and GRU-tFP} in capturing long-distance dependencies, especially when features are flattened, weakening the relationship between nearby tokens. In contrast, \sol performs well in understanding long sequences, which is important for identifying both Bot and Infiltration. These attacks often exhibit subtle, long-range dependencies in their behavior patterns that simpler models struggle to capture.

\begin{table}[H]
\centering
\caption{Performance metrics for attack detection~(Task 2)}
\resizebox{0.96\columnwidth}{!}{%
\begin{tabular}{lcccccc}
\toprule
 \textbf{Model Type} & \textbf{Accuracy} & \textbf{Precision} & \textbf{Recall} & \textbf{F1-Score} & FPR & {Inference Time (us)}\\
\midrule
CD-LSTM~\cite{LSTM-baseline-2019-IM} &  0.9888 & 0.9849 & 0.9946 & 0.9898 &  0.0182 & 4.0 \\
GRU-tFP~\cite{Usenix-2023-GRU-CNN} & 0.9839 & 0.9771 & 0.9937 & 0.9854 & 0.0279 & 1.9 \\
\midrule
\midrule
\sol & \textbf{0.9941} & \textbf{0.9978} & \textbf{0.9893} & \textbf{0.9935}  & \textbf{0.0018} & \textbf{0.75}  \\
\bottomrule
\end{tabular}%
}

\label{tab:performance_scores_task2}
\end{table}

\underline{Inference time}: We evaluate the inference time for the different models. LSTM-NoD model exhibits the highest inference time of 4.0 \textmu{}s, whereas \sol processing sequences in parallel, achieves the lowest inference time of 0.75 \textmu{}s (see Table~\ref{tab:performance_scores_task2}).

\subsection{\textbf{Task 3: Multi-class Device Classification}}\label{case_study-3-iot}

\noindent{\bf Threat model:} 
In IoT device classification (Task~3), the goal of the system, in this case a network defense system, is to identify the types of devices connected to the network by continuously monitoring its traffic flows, such as those in an enterprise environment. This helps the enterprise maintain awareness of all devices on its network and take action against unauthorized or rogue devices.

\noindent{\bf Dataset:} We utilize the UNSW 2018 dataset~\cite{UNSW-2018}, which encompasses a diverse array of device types (28 devices) exhibiting heterogeneous traffic patterns. To mitigate skewed data distributions, we train a multi-class classification head on a balanced subset of 15 selected device categories from the original 28, leveraging cross-entropy loss to enhance classification boundaries. Considering the dataset does not have labels, we group the traffic by MAC address based on the device name list. Since the dataset is imbalanced, we remove devices with very few data points and select 15 devices with more than 10,000 data points. For devices with an excessive number of data points, we randomly select 60,000 data points for each device type.

\noindent{\bf Input representation:} In this session-level task, the data is represented as sessions, where packets grouped by a src (dst) IP address within a static time-window form a session (refer Section~\ref{sec:t-mat-def}). Each session may contain multiple flows; and a single flow may span multiple sessions, thereby becoming incomplete in session(s) due to the time-window splits. The data is then segmented them into sequences of 2,000 tokens based on their arrival time. The segment labels for \sol are `0's for incomplete flow-level features and `1's for per-packet features. As for \sol~w/o~\rep, the segment labels are set to all `1's. Positional information is based on the arrival time of each packet. We use only the six default packet features mentioned in Section~\ref{sec:t-mat-def}: source/destination port representation, direction, packet size, transport layer protocol, and IAT.


\noindent{\bf Baselines:} We compare \sol with two recent sequence models for IoT fingerprinting: SANE~\cite{WU-ZEST} and BiLSTM-iFP~\cite{IoT-biLSTM-2020}. The SANE model employs a similar architecture to \sol, utilizing an attention-based structure but relying solely on per-packet features for IoT fingerprinting. Moreover, each packet is treated as a token in SANE; while in \sol, each feature is treated as a token. The BiLSTM-iFP model extracts packet-level features and uses an enhanced bidirectional LSTM to perform device classification. 

Both baseline models were implemented using single-level representations. Additionally, to analyze the impact of \rep, we conduct an ablation study comparing the performance of \sol with and without \rep~(\sol-w/o-\rep). \hlt{Moreover, given the imbalanced in the dataset, there is a risk that classes with fewer data points, i.e., \textit{minority classes}, may be overlooked or underrepresented in model training. To assess this, we specifically study the performance of four classes with the least number of data points: i)~Android Phone, ii)~Light Bulbs LiFX Smart Bulb, iii)~Smart Baby Monitor, and iv)~Aura Smart Sleep Sensor. A good performance on these classes would indicate that the model is not biased towards classes with larger data representation, thereby ensuring a more robust system.} 

\begin{figure}[ht]
    \centering
    \includegraphics[width=0.9\linewidth]{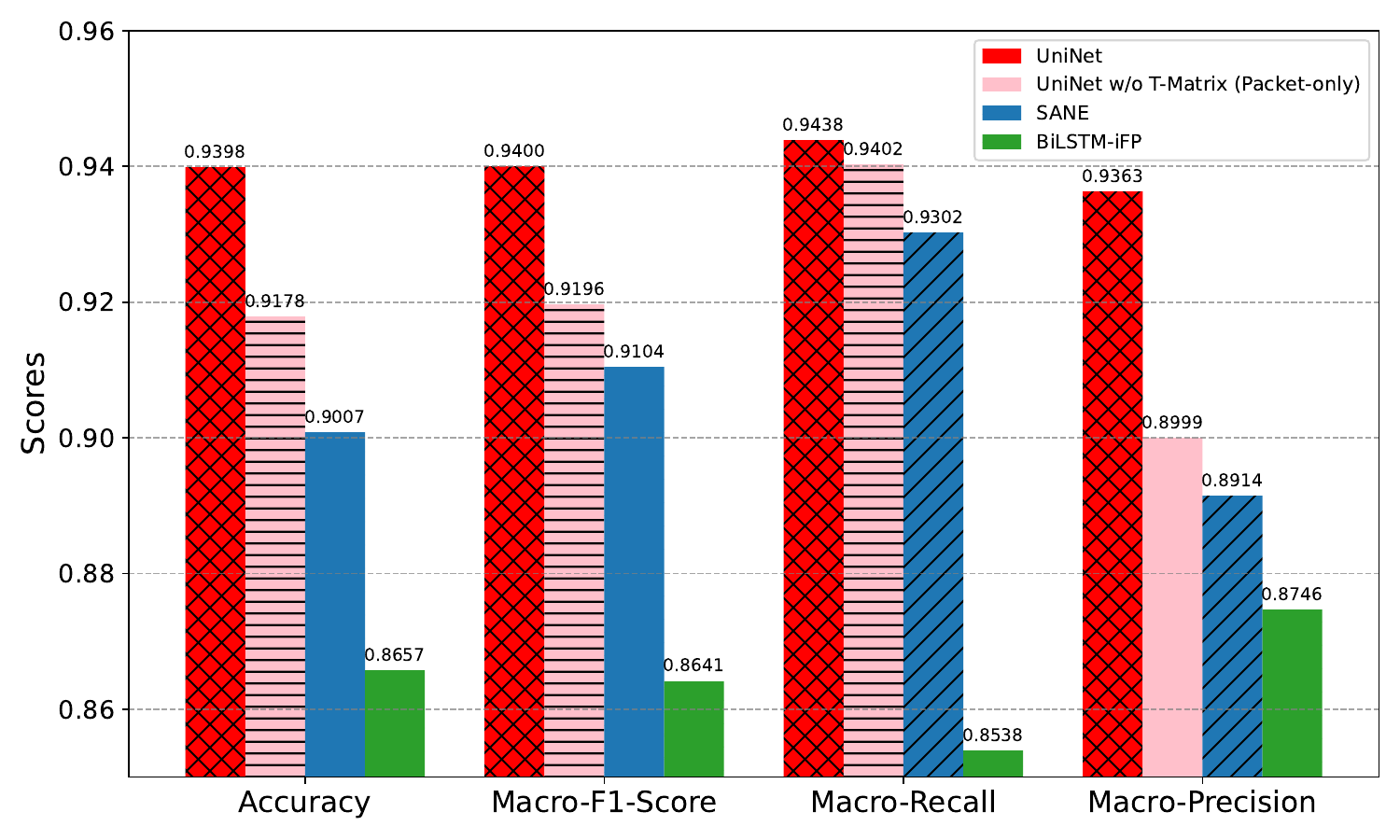} 
    \caption{{Performance comparison of \textit{minority classes} for Task~4} }
    \label{fig:iotresult-minority}
\end{figure}

\begin{table*}[!ht]
\centering
\caption{\hlt{Performance metrics comparison across overall data and minority classes. ``\sol~w/o~\rep'' refers to \sol without \rep, using a single-level representation as baselines for Task~4.}}
\resizebox{1.0\textwidth}{!}{%
\begin{tabular}{lcccccccccc}
\toprule
\multirow{2}{*}{\textbf{Methods}} & \multicolumn{4}{c}{\textbf{Overall Performance}} & \multicolumn{4}{c}{\textbf{Minority Classes Performance}} & {\textbf{Inference Time (\textmu{}s)}} \\
\cmidrule(lr){2-5} \cmidrule(lr){6-9} \cmidrule(lr){10-10}
& \textbf{Accuracy} & \textbf{Macro-Precision} & \textbf{Macro-Recall} & \textbf{Macro-F1-Score} & \textbf{Accuracy} & \textbf{Macro-F1-Score} & \textbf{Recall} & \textbf{Precision} & {\textbf{}} \\
\midrule
SANE~\cite{WU-ZEST} & 0.9841 & 0.9720 & 0.9830 & 0.9775 & 0.9007 & 0.9104 & 0.9302 & 0.8914 & \textbf{0.72} \\
BiLSTM-iFP~\cite{IoT-biLSTM-2020} & 0.9752 & 0.9514 & 0.9598 & 0.9556 & 0.8657 & 0.8641 & 0.8538 & 0.8746 & 5.90 \\
\midrule
\midrule
\hlt{\sol~w/o~\rep (session only)} & \hlt{0.6213} & \hlt{0.5897} & \hlt{0.6114} & \hlt{0.5789} & \hlt{0.5721} & \hlt{0.4836} & \hlt{0.4712} & \hlt{0.4893} & \hlt{0.65} \\
\hlt{\sol~w/o~\rep (flow only)} & \hlt{0.8125} & \hlt{0.8050} & \hlt{0.7820} & \hlt{0.7934} & \hlt{0.7581} & \hlt{0.7649} & \hlt{0.7721} & \hlt{0.7583} & \hlt{0.71} \\
{\sol~w/o~\rep (packet only)} & 0.9856 & 0.9774 & 0.9811 & 0.9792 & 0.9178 & 0.9196 & 0.9402 & 0.8999 & 0.83 \\
{\sol} & \textbf{0.9901} & \textbf{0.9886} & \textbf{0.9855} & \textbf{0.9871} & \textbf{0.9398} & \textbf{0.9400} & \textbf{0.9438} & \textbf{0.9363} & 0.85 \\
\bottomrule
\end{tabular}%
}

\label{tab:performance_metrics_combined}
\end{table*}
\noindent{\bf Orchestration of \sol:} 
Our framework addresses the challenge of incomplete flows by aggregating traffic at the session level. This preserves essential contextual relationships, enabling the detection of inconsistencies in traffic behavior that may indicate adversarial manipulation.

\noindent{\bf Analysis:} We focus on the performance of different methods on \textit{minority classes}, presented in Figure~\ref{fig:iotresult-minority}. \sol achieves the best performance across all metrics, with an improvement of $\sim 7\%$ in accuracy, $\sim 8\%$ in F1-score, and $\sim 6\%$ in precision compared with BiLSTM-iFP. 
We carry out further analyses. \hlt{i) To evaluate the advantages of the \rep, we conduct an ablation study comparing \sol with and without the multi-level representation. The versions of \textit{\sol-w/o-\rep} are divided into three categories, each using only one level of input: session-level, flow-level, or packet-level features. To ensure a fair comparison, session-level and flow-level features are positioned at the start of the input sequence and padded to a fixed length, limiting the number of flows that can be represented.
As shown in Figure~\ref{fig:iotresult-minority} and Table~\ref{tab:performance_metrics_combined}, \sol consistently outperforms all single-level input variants by a substantial margin. Compared to the session-only model, \sol improves overall accuracy and minority-class accuracy by {$\sim$60\%}, and macro-F1 score by {$\sim$90\%}. When compared to the flow-only model, \sol achieves a relative improvement of {$\sim$20\%} in overall accuracy, minority-class accuracy, and macro-F1. Even against the strong packet-only baseline, \sol delivers additional gains---improving, minority-class accuracy by {2.4\%}, and macro-F1 by {2.2\%}. These results highlight that while packet-level features are crucial, the multi-granularity integration through \rep leads to significantly more robust and generalizable performance, particularly for underrepresented classes.
}

~ii) As for the effectiveness of \model, we compare the performance between \textit{\sol-w/o-\rep (packet only)} and SANE. \hlt{Both models use advanced attention-based architectures and single-level representations. 
The key difference lies in their tokenization mechanisms: \textit{\sol-w/o-\rep} takes a feature as a token, whereas SANE is based on per-packet tokens. We observe a modest improvement in accuracy. By analyzing interactions between flows and packets within a session, and combining flow-level and packet-level features, \sol generates robust device identification.} This makes it significantly harder for adversaries to impersonate a targeted device class or maintain consistent false signals across multiple flows. The overall performance of different methods of device classification is summarized in Table~\ref{tab:performance_metrics_combined}.

\underline{Inference time}:   Table~\ref{tab:performance_metrics_combined} also provides the inference time for the different models. While BiLSTM-iFP takes 5.9 \textmu{}s, \sol, with an inference time of 0.85 \textmu{}s, is significantly faster, making it a better candidate for deployments.

\subsection{\textbf{Task 4: Encrypted website fingerprinting}}\label{casestudy-4-website}

\noindent\textbf{Threat model:} In website fingerprinting (Task 4), an adversary aims to infer which website a user is visiting based on observed traffic patterns, even when packet payloads are encrypted. We assume the attacker has a vantage point to observe client communication (e.g., compromised router) and sufficient knowledge to inspect flow and session-level characteristics, particularly in HTTP/3 (QUIC) and DNS-over-QUIC (DoQ) traffic. In the {\em closed-world} setting, the user activities are restricted to a known, ``monitored” set of websites, each of which the attacker has previously profiled through multiple training samples. Here, the adversary’s objective is to classify which monitored site the user is visiting. In the {\em open-world} setting, the users also visit an extensive set of ``unmonitored” sites. The attacker thus seeks to determine whether a given visit is to one of the monitored sites, or to an unmonitored one, despite incomplete knowledge of these unknown destinations.

\noindent\textbf{Dataset:} We use the recent DoQ-2024~\cite{DoQ-2024-dataset}, which captures network traffic from HTTP/3 and DoQ web sessions across four vantage points. The dataset includes over 75,000 unique websites, with 500 monitored QUIC sites visited 1,280 times each, and additional unmonitored sites visited 4 times each.

\noindent\textbf{Input representation:} This session-based collection allows us to extract aggregated session-level features, including the total number of flows, average and standard deviation of flow sizes and durations, total inbound and outbound bytes, and the inbound/outbound traffic ratio. These eight session-level features are concatenated with 1,992 packet-level features to form a 2,000-dimensional input vector. In our \sol architecture, we incorporate a relative embedding to distinguish session-level from packet-level segments, ensuring effective attention across both granularity.

\noindent{\bf Baselines:} We evaluate our method against several baselines, including models introduced in related works. Specifically, we compare our approach to an AutoWFP model~\cite{NDSS-2018-wfp-lstm}, the TMWF model~\cite{2023-Transformer-WFP-baseline}, and TDoQ model~\cite{Levi-2024}. AutoWFP is based on LSTM. \hlt{Although TMWF and TDoQ are based on transformer, their architectures differ significantly. TMWF employs a traditional transformer by Vaswani et al.~\cite{vaswani2017attention}, while TDoQ model utilizes a ViT-based patch embedding design~\cite{VIT}. \sol further distinguishes itself by incorporating a multi-granularity representation, \rep, combining session-level features with packet-level details, along with an expanded and more sophisticated encoding strategy (refer Section~\ref{sec:encoding})}.

\noindent{\bf Orchestration of \sol:}  QUIC/DoQ encryption conceals packet payloads, but does not entirely mask metadata such as flow sizes, inter-arrival times, and directionality, enabling the attacker to extract session-level aggregates (e.g., total flows) and packet-level features for fingerprinting. By constructing a robust signature from these features, the attacker attempts to discriminate among thousands of potential websites in both closed-world and open-world environments.

\noindent{\bf Analysis:} In our \textit{closed-world} experiments involving 300 monitored websites, we evaluate four fingerprinting methods using metrics such as accuracy, macro-precision, and F1 score. As shown in Table~\ref{tab:closed_world_metrics_refined}, \sol achieves an accuracy of 98.9\%, representing an absolute improvement of approximately 2\% over the next best method, TDoQ (96.8\%). Furthermore, \sol enhances macro-precision and F1 score by approximately 3\% each compared to TDoQ. These substantial improvements demonstrate that the multi-granular transformer architecture of \sol significantly outperforms baseline methods, thereby establishing a new benchmark in closed-world website fingerprinting.

\begin{table}[ht]
\centering
\caption{Performance of closed-world setting (300 Classes)}
\resizebox{0.95\columnwidth}{!}{%
\begin{tabular}{lcccc}
\hline
\textbf{Method}       & \textbf{Accuracy (\%)} & \textbf{Macro-Precision (\%)}  & \textbf{Macro-F1 Score (\%)} \\ \hline
AutoWFP              & 91.1                  & 89.5                                 & 89.8                  \\
TMWF                 & 92.9                  & 91.0                                  & 91.5                  \\
TDoQ           & 96.8                  & 95.0                              & 95.7                  \\
\hline
\textbf{UniNet}      & \textbf{98.9}         & \textbf{98.3}               & \textbf{98.6}         \\ \hline
\end{tabular}%
}

\label{tab:closed_world_metrics_refined}
\end{table}

\hlt{Open-world website fingerprinting: To evaluate \sol's performance in a realistic {\em open-world} scenario, we consider the top 100 QUIC-enabled domains, each generating 360 traces (36,000 traces in total) as ``monitored'', and assigned them to 100 distinct classes. An unmonitored class comprised 45,000 other websites, each contributing four traces, resulting in 180,000 traces.  Importantly, no unmonitored website appears in both the training and test sets. As per~\cite{DoQ-2024-dataset}, traces were randomly collected from various locations to ensure diversity. We employ a 75:25 train–test split for the monitored classes and a balanced 1:1 split for the unmonitored class. This features a highly imbalanced testing ratio of approximately \textbf{1:10} between monitored and unmonitored traces. We assess the TPR against the FPR in detecting monitored sites.} As is common in literature~(e.g., \cite{NDSS-2018-wfp-lstm, CCS-2023-website-fingerprint,2023-Transformer-WFP-baseline}), we adopt a binary setting by aggregating all monitored classes into a single positive category and all unmonitored classes into a single negative category.

\begin{figure}[htbp]
    \centering
    \includegraphics[width=0.95\linewidth]{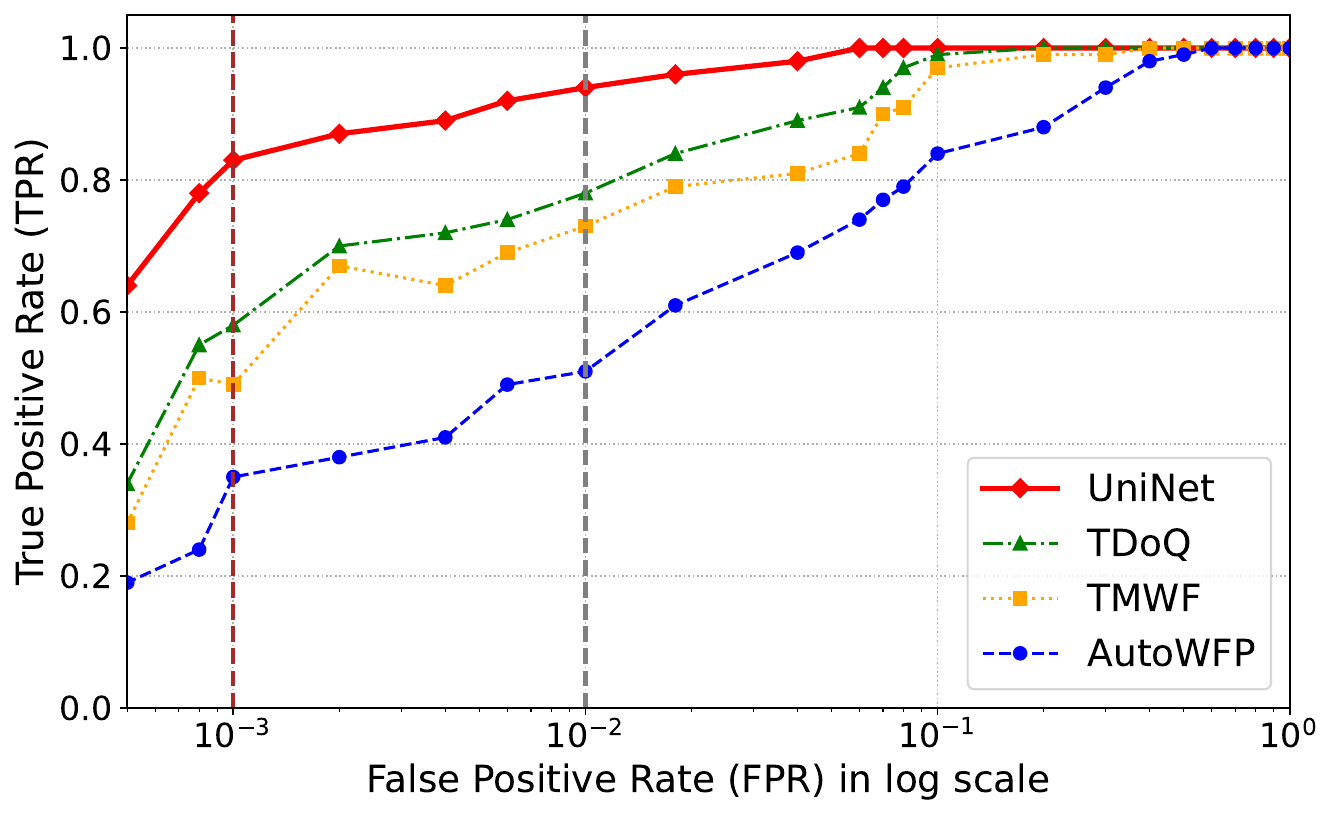} 
    \caption{Performance of open-world website fingerprinting}
    \label{fig:WFP-results}
\end{figure}

As depicted in Figure~\ref{fig:WFP-results}, \sol achieves a higher TPR at low FPR levels compared to baseline methods, demonstrating superior discriminative capabilities between monitored and unmonitored traffic. Notably, \sol attains a TPR of 81\% at a low FPR of $10^{-3}$, surpassing TDoQ (58\%), TMWF (49\%), and AutoWFP (35\%). High TPRs at low FPRs indicate that \sol can accurately identify monitored websites while maintaining a low rate of misclassification for unmonitored websites. 

\underline{Inference time}: \sol achieves the lowest average inference time of 0.15 \textmu{}s, close to that of TDoQ (0.16 \textmu{}s) and approximately one-third of TMWF's (0.45 \textmu{}s), while being just  $\approx 3\%$ of AutoWFP (4.83 \textmu{}s).

\section{Discussions and future works}\label{sec:discussion}
We now discuss the practical considerations regarding the implementation and deployment of \sol.

\noindent{\bf Model complexity and running time:} For most tasks (Task 2-4), we utilize a lightweight hierarchical transformer architecture, achieving a training time of approximately 30 seconds per epoch with a batch size of 64 samples. This demonstrates the efficiency of training. The inference time analysis shows that \sol achieves shorter inference time compared to DL baselines. For Task 1, which focuses on representation learning for traffic understanding, the model requires more data and time to train. However, this investment benefits deployment, as the pre-trained representation accelerates convergence in downstream models, ensuring overall efficiency in practical applications.

\noindent{\bf False alarm rate:} We emphasize the importance of controlling false alarms, as real-world deployment necessitates low false positive rates to reduce the operational burden on network administrators. Through our evaluations of FPR vs. TPR across multiple tasks, we demonstrate the effectiveness of \sol in maintaining a low false positive rate, making it a practical and reliable choice for network security applications.

Looking ahead, there are opportunities to enhance the architecture and expand its capabilities. 

\noindent{\bf Explainable AI (XAI) solutions:} While \sol excels in extracting contextual relationships through its attention mechanisms, its reliance on these techniques poses interpretability challenges. As a next step, we plan to incorporate XAI solutions, such as attention visualization and feature attribution, to enhance transparency and enable analysts to validate decisions. However, current XAI techniques for transformer-based models, such as gradient-based~\cite{XAI-2022-LRP-ICML}, attention-score-based~\cite{XAI-roll-out-2020}, or hybrid methods~\cite{XAI-hybrid-transformer-cvpr-2022}, are still in the early stages of development and yet to be adopted. This gap presents an ongoing challenge that we are actively exploring.

\noindent{\bf Robustness against generative evasion attacks:} We will evaluate the robustness of our \sol against adversarial attacks. In particular, we assess its resilience to evasion techniques—a critical issue in traffic analysis. Attackers may use methods such as traffic manipulation, adversarial perturbations, or obfuscation to circumvent machine learning-based traffic analysis systems.

\section{Related works}\label{sec: related works}

Below, we discuss three critical stages in the ML-based traffic analysis pipeline: feature representation, feature encoding, and model development. By examining current approaches at each stage, we identify trade-offs that underscore the need for a unified, more adaptive framework.

\subsection{Feature representation}\label{feature-rep-relatedwork}

Existing feature representation techniques fall mainly into two categories: \textit{bit-level} and \textit{semantic} representations. {\em Bit-level} representation uses the raw binary bits from the packet header to represent each packet~\cite{ nprint2021, bit-level-2022-KDD, bit-level-2024-trans}. 
This method can be enhanced to ensure field alignment between packets of different protocols, e.g., using padding~\cite{nprint2021}. Since the header info of each packet is encoded using bit values, this is a per-packet representation. However, such a simple encoding has two serious limitations: 
\begin{itemize}
    \item Bit-level representation of header hard codes certain fields, such as src/dst IP address, leading to model overfitting. For example, in most cases, a benign computer that is infected or breached may start communicating with a C\&C server. However, if the model has seen only benign traffic from this IP address, then it would likely classify the attack flow as malicious because of overfitting the IP address. nPrint proposed in~\cite{nprint2021} exhibits this overfitting tendency as the results are dependent on attacker IP addresses. Similarly, due to randomness, encoding ephemeral ports as such is not useful and might mislead a model.

    \item When using bit-level features for unsupervised representation learning, the smallest token unit is typically one byte (e.g., as in~\cite{NetGPT}). This approach can disrupt meaningful fields due to the varying field sizes. For example, the 16-bit port number in the header would be split into two tokens instead of being represented by a single token. Furthermore, bit-level representation increases the model size when provided as input to a sequence model, leading to a higher consumption of resources (compute and memory), besides increasing the inference time. For instance, a header with a minimum of 20 bytes would require at least 20 tokens to represent a single packet.
\end{itemize}

{\em Semantic} representations typically aggregate multiple packets or flows into constant-size feature vectors. For instance, repeated failed connection attempts to diverse destinations can signify bot activity reaching out to command-and-control (C\&C) servers. Aggregated features are widely used in network security tasks, such as anomaly/attack detection~\cite{Usenix-2021-sequence-anomaly}, botnet detection~\cite{botnet-S&P-2020} fingerprinting~\cite{CCS-2023-website-fingerprint}, etc. While semantic features can capture meaningful higher-level indicators (e.g., port usage, flow durations), they rely heavily on domain expertise. This makes them less flexible in scenarios with limited or evolving domain knowledge.

\subsection{Feature encoding for ML training}

Feature encoding transforms network traffic data into numeric representations suitable for ML models~\cite{GAN-traffic-generation-sp-2020}. The process begins with normalizing heterogeneous data into a unified format, ensuring consistency and facilitating effective encoding. After normalization, data is tokenized into its minimum units for fine-grained analysis. These tokens are then embedded to extract relationships essential for understanding network behaviors. However, existing encoding methods often fall short in practical network traffic analysis~\cite{ET-BERT,NetGPT}. For instance, one-hot encoding, commonly used for categorical features like port numbers, creates high-dimensional sparse vectors, thereby increasing the computational complexity and the risk of overfitting~\cite{GAN-traffic-generation-sp-2020}. Embedding techniques like Word2Vec~\cite{word2vec} have been adopted in NLP, with newer contextual embedding methods proving more effective~\cite{Roberta-2019}. However, current approaches often use raw hex numbers for tokens~\cite{NetGPT,ET-BERT,peng2024ptu,lens-arXiv-2024}, which fragment fields into less meaningful pieces. Treating entire packets as single tokens has been proposed but poses challenges due to high dimensionality, leading to large vocabularies that complicate training~\cite{HotNets-rethinking}. Additionally, most of these works overlook sequential information between packets, such as inter-arrival time (IAT), which is helpful in capturing temporal patterns in network traffic.

\subsection{Models for network traffic analysis}

A wide range of models have been developed for analyzing network traffic. In~\cite{NTU-2025-Survey} and the works it surveys, statistical methods, ML, and DL models have been widely applied to tasks such as anomaly detection, device and website fingerprinting, location inference, quality of experience (QoE) measurement, and traffic classification. Statistical models rely on well-established statistical principles to identify anomalies or deviations from normal traffic patterns~\cite{infocom-2009-Statistic,2011-ICC-Statistic, anomaly-detection-alpha-stable-2011}. However, they often struggle with complex, evolving threats, as they rely on predefined statistical assumptions that attackers can circumvent. ML models offer greater flexibility by being able to learn from data. Techniques such as decision trees, support vector machines (SVM), and ensemble methods like Random Forests have been widely used to classify network traffic, detect intrusions, manage resources, fingerprint IoT devices, etc.~\cite{one-classSVM-2003,kmeans-2008-anomaly,isolation-forest-2018,LOF-2019,D3T-2024,feng2023explainable-reviewers, DEFT-2019, ADEPT-2021}. These models can adapt to new data, improving detection rates over time. However, they often require significant feature engineering and may struggle with the high dimensionality of network data. DL models, including CNNs, recurrent neural networks~(RNNs), and transformers, are capable of extracting meaningful information from raw data, capturing sequential patterns and relationships that traditional ML models might miss~\cite{GEE-2019, NDSS-2020-flowprint,NPRA-2023, Usenix-2023-GRU-CNN}. DL models are also particularly good at handling large-scale data and can potentially adapt to various types of threats and attacks~\cite{GEE-2019}. Nevertheless, they require substantial computational resources and large labeled datasets for training and model maintenance, which can be a barrier to their widespread adoption. A common disadvantage of the current solutions is that they often rely on task-specific models, which may not generalize well across different types of network anomalies or attack vectors~\cite{HotNets-rethinking,NetSPAI-2025}.

\section{Conclusions and Future Work}\label{sec:conclusion}

In this work, we presented \sol, a unified framework for network traffic analysis that introduces the \rep multi-granularity representation and the lightweight attention-based model, \model. \sol addresses key limitations of existing approaches by seamlessly integrating session-level, flow-level, and packet-level features, enabling comprehensive contextual understanding of network behavior. Its adaptable architecture, featuring task-specific heads, supports a variety of network security tasks, including anomaly detection, attack classification, IoT device fingerprinting, and encrypted website fingerprinting. Extensive evaluations across diverse datasets demonstrated the superiority of \sol over state-of-the-art methods in terms of accuracy, false positive rates, scalability, and computational efficiency.

In recent years, the networking community is exploring ways to build network foundation models so as to apply them to multiple downstream tasks across different network environments. Representation is a key aspect of a foundation model~\cite{NetSPAI-2025}; a common approach for representation encodes raw packet bytes as hex‐value tokens~\cite{NetGPT,ET-BERT,peng2024ptu,lens-arXiv-2024}, but this byte‐level tokenization fragments protocol fields and obscures high‐level semantics~\cite{HotNets-rethinking}. \sol decomposes each packet into coherent units---headers, options, payload---to reduce vocabulary size while preserving semantic structure, then augments each unit embedding with inter-arrival times to capture temporal dynamics. A hierarchical transformer first models per-packet semantics and timing, then captures cross-packet dependencies, yielding efficient training and robust generalization across tasks such as anomaly detection, performance prediction, and traffic classification. This may bring us one step closer to a powerful network foundation model.

Another direction worth exploring is traffic generation, which could enhance system robustness by generating or augmenting missing or synthetic data points. Leveraging the learned representations from \sol, we could integrate generative models such as auto-encoders, Generative Adversarial Networks (GANs), diffusion models, or transformer decoders to generate network traffic~\cite{Netdifussion-2024-traffic-generation, GAN-traffic-generation-sp-2020,Sigcom2022-generative,2024-NDSS-data-augmentation}. This would help address challenges due to privacy and data sparsity, while improving model reliability in diverse scenarios. This area of research is extensive and beyond the scope of the current work, leaving it as an exciting opportunity for future exploration.

\bibliographystyle{IEEEtran}
\bibliography{document}

\end{document}